\documentclass[twocolumn]{aastex631}

\begin{document}

\title{A high-resolution imaging survey of massive young stellar objects in the Magellanic Clouds}

\author{Venu M. Kalari}
\affiliation{Gemini Observatory/NSF’s NOIRLab, Casilla 603, La Serena, Chile}

\author{Ricardo Salinas}
\affiliation{Departamento de Astronom\'ia, Universidad de La Serena, Av. Juan Cisternas 1200, La Serena, Chile}

\author{Hans Zinnecker}
\affiliation{Universidad Autonoma de Chile, Pedro de Valdivia 425, Providencia, Santiago de Chile, Chile}

\author{Monica Rubio}
\affiliation{Departamento de Astronom\'ia, Universidad de Chile, Santiago, Chile}

\author{Gregory Herczeg}
\affiliation{Kavli Institute for Astronomy and Astrophysics, Peking University, Yiheyuan 5, Haidian Qu, 100871 Beĳing, People’s Republic of China}
\affiliation{Department of Astronomy, Peking University, No.5 Yiheyuan Road, Haidian District, Beĳing 100871, People’s Republic of China}

\author{Morten Andersen}
\affiliation{European Southern Observatory, Karl-Schwarzschild-Strasse 2,
85748 Garching bei München, Germany}




\begin{abstract}

Constraints on the binary fraction of young massive stellar objects (mYSOs) are important for binary and massive star formation theory. Here, we present speckle imaging of 34 mYSOs located in the Large (1/2\,$Z_{\odot}$) and Small Magellanic Clouds ($\sim$1/5\,$Z_{\odot}$), probing projected separations between the 2000--20000\,au (at angular scales of 0.02-0.2$\arcsec$) range, for stars above 8\,$M_{\odot}$. We find two wide binaries in the Large Magellanic Cloud (from a sample of 23 targets), but none in a sample of 11 in the Small Magellanic Cloud, leading us to adopt a wide binary fraction of 9$\pm$5\%, and $<$5\%, respectively. We rule out a wide binary fraction greater than 35\% in the Large, and 38\% in the Small Magellanic Cloud at the 99\% confidence level. This is in contrast to the wide binary fraction of mYSOs in the Milky Way (presumed $Z_{\odot}$), which within the physical parameter space probed by this study is $\sim$15-60\% from the literature. We argue that while selection effects could be responsible for the lower binary fraction observed; it is more likely that there are underlying physical mechanisms responsible for the observed properties. This indicates that metallicity and environmental effects may influence the formation of wide binaries among massive stars. Future larger, statistically more significant samples of high-mass systems in low-metallicity environments, and for comparison in the Milky Way, are essential to confirm or repudiate our claim.

\end{abstract}



\section{Introduction} \label{sec:intro}

Observations of massive stars suggest that most of them are found in multiples \citep{sana14}. Often, they come as higher order multiples, with close spectroscopic binaries and wider visual companions, in the range of $\sim$10s--10000s of au. The formation of such systems with physical separations less than 0.1\,pc (or 20000\,au, a crude limit for physically bound systems) is thought to take place during the earliest stage of star formation \cite{offner}. To understand the formation of these systems, one needs to distinguish between primordial multiplicity (for e.g. due to a metallicity effect on radiative feedback), and secondary evolutionary effects on multiplicity (e.g. N-body dynamics).


\begin{figure*}
\plottwo{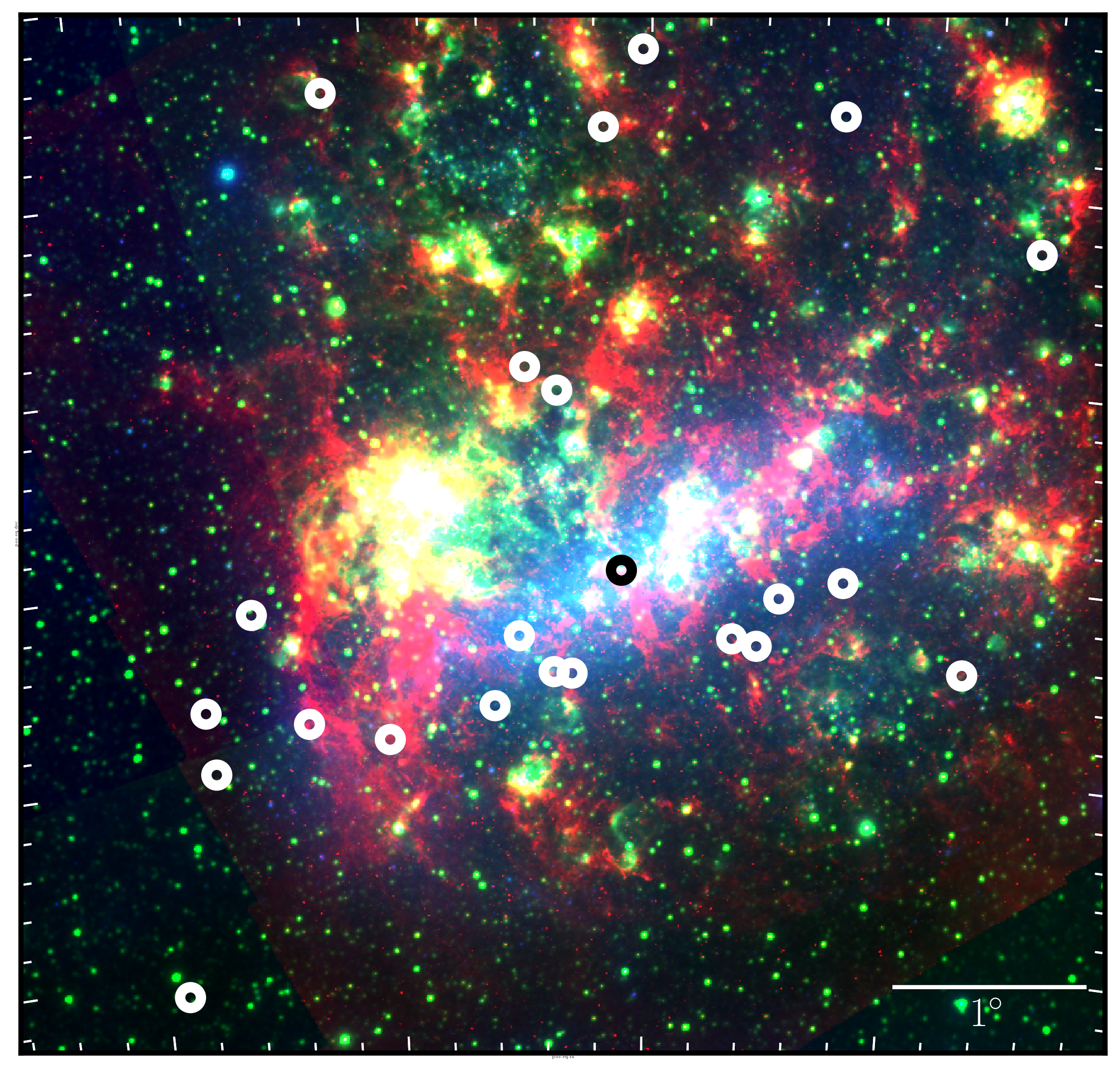}{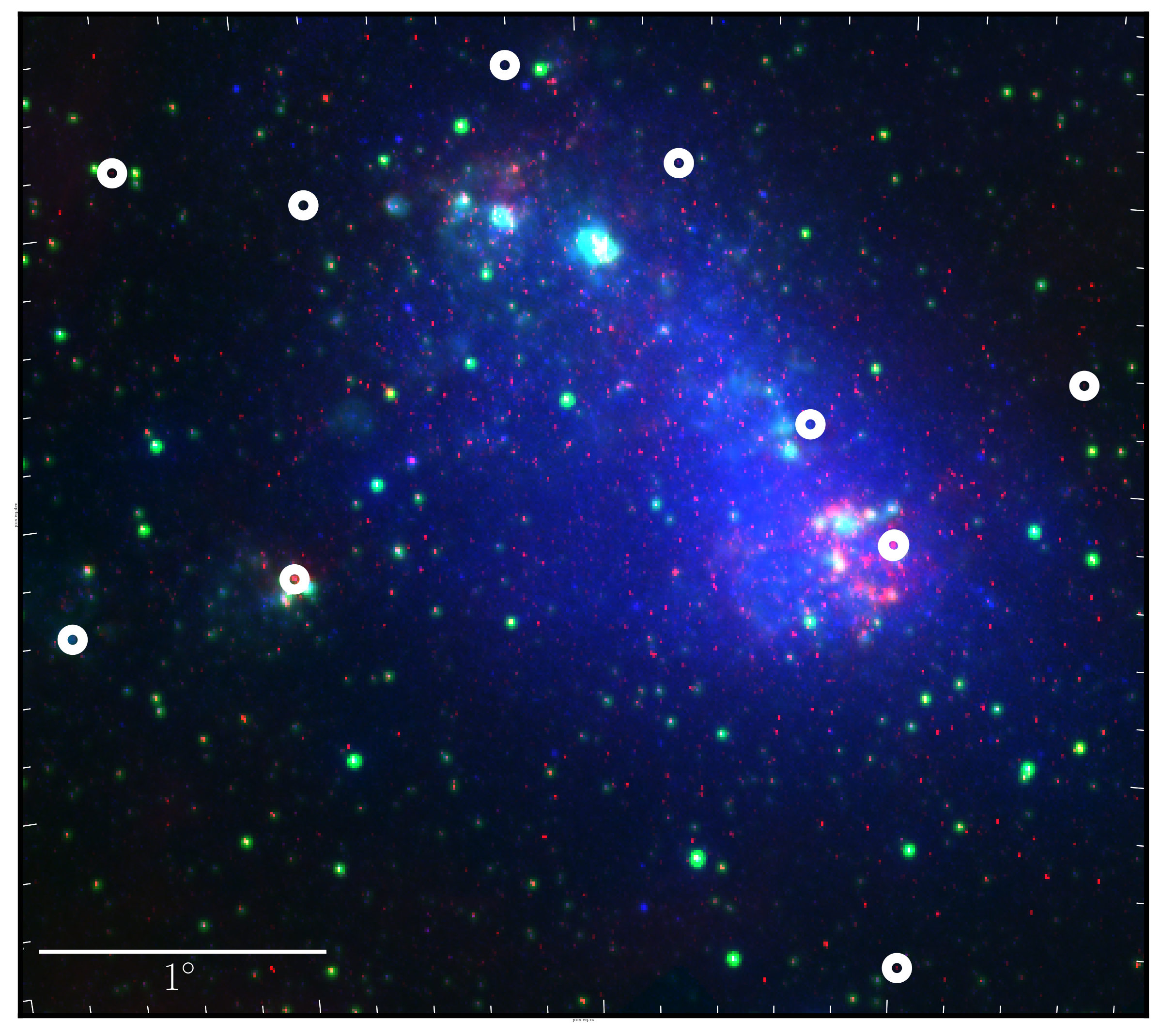}
\caption{{\it rgb} image of the LMC ({\it left}) and SMC ({\it right}), where the red represents {\it Spitzer} 8$\mu$m emission \citep{whitney, sewilo13}, green H$\alpha$ \citep{haimage}, and blue the {\it DSS} blue filter. The LMC image is centered on 5$^h19^m45^s$, $-68^d48^m47^s$ over a 5.6$^\circ \times 5.3^\circ$ field, and the SMC image is centered on 1$^h$00$^m$15$s$, $-73^d06^m05^s$ covering a 3.9$^\circ \times 3.5^\circ$ area. White circles mark the location of the observed mYSOs, with the scale bar given in the bottom corner. North is up and east is to the left.    \label{smcpos}}
\end{figure*}


Main-sequence massive star binaries may, or may not have been born in such configurations. Massive young stellar objects (mYSOs) are stars in the very early stage of formation ($\lesssim$\,1\,Myr), and the progenitors of massive main sequence stars. They are sufficiently young that secondary evolutionary effects do not influence the observed configuration. Studies of mYSOs in our Galaxy have found a wide binary fraction between 30--60\%, in the separation ranges between 1000-100000\,au \citep{pomohaci, shenton, bordier}. 

Searching for wide companions to mYSOs in the metal-poor Large, and Small Magellanic Cloud (LMC, SMC at 1/5 and 1/3\,$Z_{\odot}$ respectively; see \citealt{russel}) at a distance of 49.5 \citep{lmcdist} and 63\,kpc \citep{smcdist}, respectively, presents a unique opportunity to explore the impacts of metallicity on binarity. Observationally, this requires very high angular resolution, to the order of 0.02--0.2$\arcsec$ to resolve companions in the 2000--20000\,au range in the Magellanic Clouds. Speckle imaging with the ZORRO instrument at Gemini South can now deliver angular resolution at such scales, reaching contrasts between 2--4\,mag at these angular scales. Magellanic cloud mYSOs in wide binaries can therefore, now be resolved. 


This opens up a new research dimension, namely to check if the multiplicity properties of mYSOs differ as a function of metallicity. Since the Magellanic Clouds have a lower extinction compared to our Galaxy, and well known distances, this allows us to create a volume, and brightness limited sample to compare to literature results in the Milky Way. In this paper, we estimate the binary fraction of mYSOs in the Magellanic Clouds observed using ZORRO imaging, and discuss the implications of our findings. Binary fraction in this work is defined as the fraction of primaries with a known companion (also referred to as the multiplicity fraction).

Our paper is organized thus. Section 2 presents our sample, while discussing the biases and caveats introduced by our selection. We also outline our observations, and the data reduction process. Our results are given in Section 3. We discuss our findings, and their implications in Section 4. The conclusions of our study, and potential future work are presented in Section 5. 

\section{Observations} \label{sec:ff}

\subsection{Sample selection}
\begin{figure*}
\includegraphics[width=1\columnwidth]{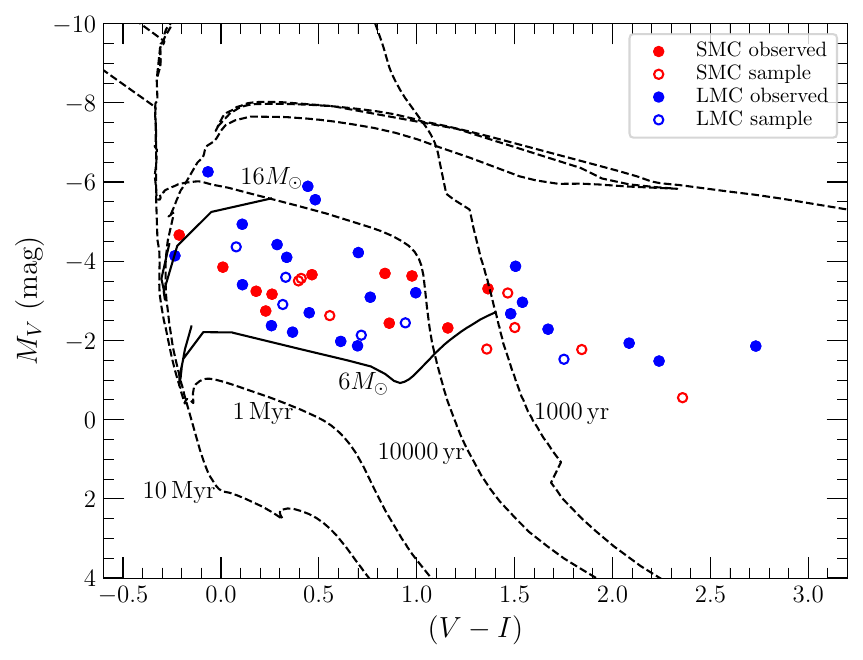}
\includegraphics[width=0.95\columnwidth]{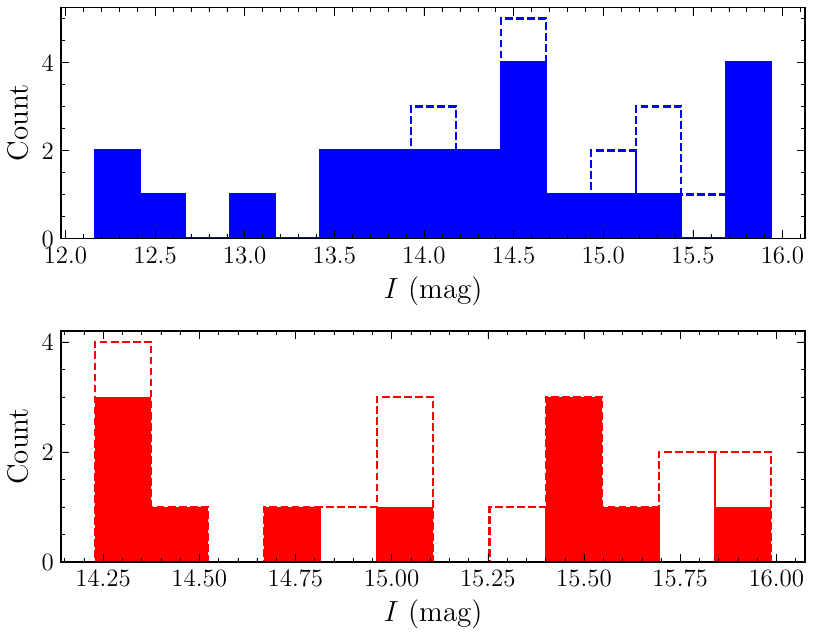}
\caption{{\it Left}: ($V-I$) vs. $V$ color-magnitude diagram of the targets. Blue (red) open circles represent the LMC (SMC) sample, while filled circles are the targets observed. Colours are not corrected for extinction. Magnitudes are corrected assuming a distance modouls of $\mu$ of 18.48 and 18.9 for the LMC and SMC respectively. The 1000\,yr, 10000\,yr, 1\,Myr, and 10\,Myr isochrones are marked by dashed lines. Solid lines represent the 6, and 16\,$M_{\odot}$ stellar mass tracks. Stellar isochrones and mass tracks are from \cite{parsec}. {\it Right}: The $I$-band histograms of the selected and observed targets for the Large Magellanic Cloud (top), and Small Magellanic Cloud (bottom). Filled histograms represent observed targets, and dotted lines represent the complete sample. \label{cmd}}
\end{figure*}

We aim to detect potential wide binaries in mYSOs in the Magellanic Clouds. To select our targets, we therefore attempt to select from a homogeneous sample of known mYSOs. mYSOs are most strongly detected in the near to mid-infrared, based on their disc/envelope excess. While one can identify them using emission line signatures, these can be confused with other more evolved emission line stars \citep{kalari14}, especially at the higher mass end, even using spectroscopy (but additionally, some mid-infrared excesses can also be confused with disc/shell bearing giant branch stars; e.g. \citealt{jones17}). 
The sample for the LMC (see Fig.\,1) consists of YSOs (young stellar objects) detected using {\it Spitzer} mid-infrared imaging at 3.6, 4.5, 5.8, and 8$\micron$ by \cite{whitney}. They applied various color cuts (driven by colors of YSO models from \citealt{rob}). They then applied further sanity checks, and visual inspection to remove contaminants such as galaxies, evolved stars, and planetary nebulae. We select from their sample only high probability mYSOs (following the classification schema in their catalog). Note that these likely do not include the most massive YSOs, since those are removed based on the stringent color cuts employed (c.f. \citealt{gchu}). {\it BVRI} magnitudes of the sample were obtained by cross-matching with \cite{mcps}, using a radius of 1$\arcsec$. $I$ magnitudes were compared with the $J$-band magnitude from \cite{whitney} to ensure objects widely discrepant in magnitude were not selected. The final target list is limited to $I<$16 mag due to the sensitivity limit of ZORRO. We note that our sample when ordered according to optical brightness creates a bias towards more evolved and massive targets.  Since all YSOs do not have equal extinction, we are biased towards objects with low extinction, and high mass following our magnitude cut. The selected objects are plotted in the color-magnitude diagram in Fig.\,\ref{cmd}. From the analysis of \cite{whitney}, our targets should have warm dust, and possibly discs/envelopes present around them. 

A similar selection procedure is followed for the SMC (see Fig.\,1). \cite{sewilo13} published a catalog of high confidence YSOs in the SMC using {\it Spitzer} mid-infrared photometry. They refined the selection criteria of \cite{whitney}. Along with using color-cuts driven by YSO models, they refined the selection using visual inspection, and model fitting to remove contaminants. They also used known YSOs detected via {\it Spitzer} spectroscopy to guide source selection criteria. Their final products includes a catalog of SMC mYSO candidates with high-reliability (following their final classification schema). Their catalog also suffers from the same issue at the higher mass end, where the highest mass sources are most likely removed due to stringent selection criteria. We apply a $I<16$ cut to their high-reliability YSO catalog to select our SMC targets.

We draw attention to the fact that the {\it Spitzer} aperture is $\sim$2$\arcsec$ (around 0.5\,pc at the distance to the Magellanic Clouds) at 3.6$\mu$m. Optical photometry from \cite{mcps, sewilo13} has a spatial resolution around 1$\arcsec$. Therefore, there is little possibility of any wide binaries being detected in our separation ranges using the current archival photometry. Higher angular resolution images for some mYSOs exist in the {\it Hubble Space Telescope} ({\it HST}) archive. The angular resolution of {\it HST} is comparable to ZORRO for objects at very close separations (see \citealt{kalari22}); offering a crucial sanity check to our study. In total, our samples contain 29 and 19 targets in the LMC and SMC respectively. The sample as a function of observed magnitude in LMC and SMC is shown in Fig.\,\ref{cmd}.

\subsubsection{Biases and caveats}
A brightness limited sample suffers from Malmquist bias. Our sample in the classical way does not, since all are objects are approximately at the same distance when viewed from Earth given the orientation of the clouds. Any small inter-object distances along our line of sight are negligible. However, we do not account for extinction. Therefore, the objects with smaller extinction values will be selected. Preferentially this would select more evolved, pre-main sequence (PMS) type objects given that evolutionarily they are more likely to have cleared their surrounding natal dust. Secondly, our selection methodology will prefer binaries. Given a single and a binary star at the same distance, same extinction and spectral type, an equal mass binary will appear twice as bright, but have similar colors. Since color criteria are used for YSO identification by \cite{whitney} and \cite{sewilo13}, more equal mass binaries (if present) than single stars will have been selected, especially towards the faint end.


Our final sample is expected to limit selection effects, since our targets are located in a variety of dense, and sparsely populated regions, and in regions of high and low extinction, and vigorous star formation (see Fig.\,1). 
Firstly, all our mYSOs lie at a uniform distance, whereas most of the Galactic known mYSOs \citep{pomohaci, shenton} lie at distances beyond 2-3kpc, where accurate and precise distances are not known, even with current {\it Gaia} \citep{gaia} parallaxes. Therefore focusing on only the Magellanic Cloud mYSOs sample, we can correctly convert the angular to linear separations, which might be subject to future changes in a Galactic sample. Secondly, the lack of dust and low Galactic latitude means that our sample suffers very low line of sight extinction compared to Galactic mYSOs, preferable in optical observations, which achieve the highest angular resolutions with ZORRO. Finally, most Galactic mYSOs are located near the Galactic plane increasing chance alignments, a problem not faced in our Magellanic Cloud sample. Regardless, we account for chance alignments statistically in our sample, relying on background source density estimates. 


\subsection{Observations} \label{sec:s}

Observations were conducted using the ZORRO speckle imager mounted on the Gemini South 8.1m telescope located at Cerro Pach{\'o}n, Chile  \citep{scott}. Observations were taken in zenith seeing of less than 0.7$\arcsec$, in clear skies, but in a variety of moon phases. Owing to the low declination of the Magellanic Clouds, the targets can never be observed as close to the meridian as possible for the best possible speckle corrections. Observations were conducted over a three year period, from September 2019-- October 2021. However, not all the targets selected were observed owing to difficulty in obtaining the observations, and scheduling constraints given that ZORRO is a visiting instrument at Gemini South (and only operates during $\sim$2--3 week long blocks in each semester).

The final observed list of mYSOs is given in Appendix\,A, along with their known magnitudes, and other identifications. They are also populated in the magnitude histograms in Fig.\,\ref{cmd}. Based on the observed sample, our LMC sample can be considered approximately complete till $I$=15, after which it is 50\% complete. The SMC sample is around 50\% observed at all magnitudes till the limit. 

\subsection{Data reduction} \label{sec:ste}

ZORRO provides simultaneous speckle imaging in two custom medium-band filters centered on 562 and 832\,nm (hereafter the 562 and 832 filters). The data are processed using the pipeline described in \cite{howell}. The pipeline calculates the power spectrum of each image, then corrects for the speckle transfer function by dividing the mean power spectrum of the target by that of the standard star. The pipeline also produces a reconstructed image of each target. Achieved angular resolutions are around 17 and 25\,mas in the 562, and 832\,nm filters, respectively. The contrast ($\Delta$m) limits for each target were determined using the method described in \cite{horch}, which uses the background flux levels to determine the faintest companions one could reliably detect at each separation. An example of reduced data products for a companion detection and non-detection is shown in Fig.\,\ref{detect}.

\section{Analysis} \label{sec:le}

We have obtained speckle imaging reaching around 17--25\,mas with contrasts ($\Delta$m) between 1--4\,mag of 23 LMC mYSOs and 11 SMC mYSOs. Our results are given below, with a discussion of the detected binaries. The complete set of reconstructed images, and contrast curves are given in Appendix\,B. 

\begin{figure*}
\plottwo{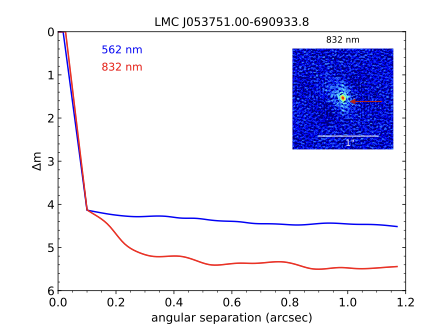}{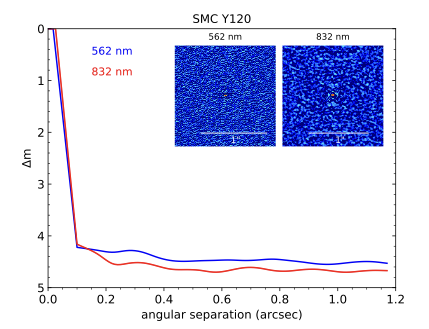}
\caption{Example of reconstructed images (inset) and contrast curves in both filters for a companion detection (left) and non-detection (right). The arrow in the left-hand figure marks the location of the detected companion. \label{detect}}
\end{figure*}

In the LMC, we observed 23 mYSOs. Of these, only three displayed evidence for a companion using speckle imaging. No higher order multiples were detected. The properties of the binary companions are given in Table\,1. In Fig.\,\ref{detect}, an example of a detection and non-detection is shown. We estimate the chance of spurious contaminants ($P_{\textrm{c}}$) to the detected binaries following \cite{pomohaci} as 1$-e^{\pi d^2 \rho}$, where $d$ is the separation, and $\rho$ the background source density. $\rho$ is calculated based on source counts from {\it Gaia} astrometry, or archival {\it HST} imaging. 

\subsection{Detected binaries}

In our sample, we detected three wide companions to known mYSOs in the LMC. LMC\,J054059.79-693840.1 is the brightest and bluest star in our sample at $V=12.24$\,mag, and ($V-I$)$=-0.06$. It is located at the north eastern edge of the LMC N160 star-forming complex, at the edge of cavity visible in {\it Spitzer} mid-infrared imaging \citep{whitney}. However, \cite{evans15} suggest that the source is a blue supergiant and a spectroscopic double lined binary based on optical spectra between 380-500\,nm, and classify it as a B1.5 Ia spectral type. The mid-infrared colors of the target would put it firmly in a YSO candidate region of color-magnitude and color-color diagrams based on various selection criteria (e.g. \citealt{gchu, sewilo13}). The companion detected by ZORRO is seen in {\it HST} imaging \citep{whitmore} at a distance of 0.69$\arcsec$. It exhibits a $\Delta$m in $I$ of $-$3.3\,mag. Both the separation and contrast are similar to the result obtained by our imaging, suggesting they are the same companion. 
We suggest that the primary is the early B supergiant detected by \cite{evans15}. The effective temperature ($T_{\rm eff}\sim$20550\,K) and luminosity (5.2\,$L_{\odot}$) based on the \cite{evans15} spectral type place the early spectral type primary well out of the range of an asymptotic giant branch (AGB) impostor, indicating that it is not the source of the exhibited mid-infrared colors. However, the mid-infrared colors (the mid-infrared imaging aperture will include both the primary and secondary) do not resemble an early B supergiant (c.f. \citealt{bonanos}). This indicates that the secondary is the most likely source of the mid-infrared colours. It is possible that the early B supergiant carved out the cavity visible in the mid-infrared, which triggered further star formation including the mYSO; or the current location of the mYSO is coincidence.  The $P_{\textrm c}$=52\%, and therefore we consider this a chance alignment. 

LMC J050830.53-692237.3 has a companion detected at 0.4$\arcsec$, but at a contrast of only 0.42\,mag. It is located in a region of known mYSOs \citep{gchu,whitney}, and a known molecular cloud from the \cite{wong} survey (A100 in that paper). Follow-up optical spectroscopy by \cite{kamath} suggests that the object is a YSO, with a line of sight carbon rich star \citep {woods}. They based their analysis on the detection of a hot component with mid-infrared spectra displaying O-rich features, and cool component with C-rich features with two velocity components. No {\it HST} images exist for this object. The chance alignment probability of this source is quite low if we assume source density based {\it Gaia} astrometry (4\%), but if we assume instead source density of J054059.79-693840.1 (representative of a more denser star-forming region in the LMC), the $P_{\textrm c}$ climbs to 28\%. All of this evidence indicates that the nature of the binary companion of J050830.53-692237.3 is inconclusive, and that it either can be a spurious chance detection and a late type C-rich star, or a physically related star to the YSO exhibiting a latter spectral type. We consider this during our estimation of the binary fraction. 

J053751.00-690933.8 is the final source with a companion detected, close to the limit both in separation (0.04$\arcsec$), and contrast ($\Delta m$=2.19\,mag). It is located in the 30 Doradus star-forming region \citep{kalari18, fahrion}. It is also marked as a high probability YSO in the study of \cite{gchu}. Interestingly, the object is also marked as a late O supergiant in \cite{vfts}. Multi-filter {\it HST} photometry of \cite{http} indicates only a source 0.25$\arcsec$, but beyond the ZORRO contrast limit ($\Delta m>$5\,mag). It is likely that the source detected here is previously unreported. Given the aperture used in \cite{vfts}, it is likely that both the YSO, and the supergiant are in a physical system seen in the ZORRO imaging, and that the optical spectral features are dominated by the supergiant; while the infrared SED shape is dominated by the mYSO. This requires further follow up from high angular resolution spectroscopy to study both components visible here. The probability of chance alignment, even assuming the high density (30317 sources\,arcmin$^{-2}$) of the 30 Doradus region (from \citealt{http}) is low at $P_{\textrm c}$=4\%,  owing to the close separation. We classify this object as a physically related binary.

\begin{deluxetable}{llllll}
\tablenum{1}
\tablecaption{Properites of LMC wide binaries detected in our study}
\tablewidth{0pt}
\tablehead{
  \colhead{Name} &
  \colhead{Sep.} &
  \colhead{Sep.} & \colhead{$\Delta$m$^1$} &\colhead{Ratio} &\colhead{$P_{\textrm c}$} \\ 
 \colhead{}& \colhead{($\arcsec$)} & \colhead{(au)} & \colhead{(mag)} &  \colhead{($M_{\odot}$)} & \colhead{(\%)} 
}
\startdata
LMC\,J054059.79-693840.1  & 0.621 & 30739 &   3.89 & 0.15 & 52 
\\ 
LMC\,J050830.53-692237.3 &  0.423 & 20938  &  0.42 & 0.82 & 29 
\\
LMC\,J053751.00-690933.8  & 0.040 & 1980   &   2.19 & 0.35 & 4 
\\
\enddata
\tablecomments{$^1$ Given in the 832\,nm filter. }
\end{deluxetable}

\subsection{Binary fractions}

Estimating the binary fraction in the Magellanic Clouds is the main aim of this study. In addition to the biases introduced by our sample selection considered in Section 2, an additional issue when estimating the binary fraction is contamination. As discussed in Section 2, even with additional data such as spectroscopy in the optical and infrared, it is not always straightforward to separate mYSOs from more evolved counterparts (also seen in Section 3.1). Therefore, the estimated binary fraction needs to consider the rate of contamination in our sample. To do so, we employ spectroscopic studies classifying sources where available from the literature in the LMC. These are tabulated in Appendix\,A. 

Summarizing our results, we find comparing to the {\it Spitzer} mid-infrared spectral classification schema of \cite{woods, jones17} a contaminant fraction $\sim$40\%, consisting of AGB stars. A comparison to the optical spectroscopic sample of \cite{kamath} discriminating between evolved and YSO targets suggests a contamination rate of 35\%. Comparing the optical and mid-infrared study, we find only one discrepancy which is that of the discussed case of J050830.53-692237.3. Note that much of the contamination is restricted to the brighter end of the sample (for e.g., all cross-matched targets having spectra classified as AGB in \citealt{jones17} are brighter than $I<13.5$\,mag), suggesting that the contamination fraction is likely lower when moving towards the fainter end. However, the lack of such spectroscopic data at the fainter end and in the SMC does not allow us to test our suggestion. Overall, we consider the total contamination fraction in our sample to be 35\%. This is the value we adopt when estimating the errors on the binary fraction of our sample. 

In the Large Magellanic Cloud, we detected three wide binaries, one of which is certainly a chance detection. This suggests that our binary fraction based on the observations within the ZORRO detection limits is 9$\pm$5\% in the LMC. The error bar arises from assuming a contamination rate of 35\% in our sample of  LMC YSOs; and if we consider J050830.53-692237.3 to be a physically related companion. Following the method of \cite{binarystats}, we can rule out at the 99\% confidence level a wide binary fraction greater than 35\% in the LMC. The \cite{binarystats} exact method allows for calculation of confidence intervals for binomial distributions. In the Small Magellanic Cloud, we observed no wide binaries within our detection limit. We are therefore unable to estimate a binary fraction but assume a minimum fraction less than 5\% for the rest of the analysis, given the sample size and assumed contaminant fraction. At the 99\% confidence interval, we can rule out a wide binary fraction greater than 38\% for our SMC sample.    

A notable result of our study is that for most optically visible mYSOs identified by {\it Spitzer} in the Magellanic Clouds, the primary light dominates on scales of few arcsec, lending statistical support to the parameters derived from fitting models to lower angular resolution mid-infrared imaging (e.g. \citealt{sewilo13, kalari20}).

\begin{figure*}
\plottwo{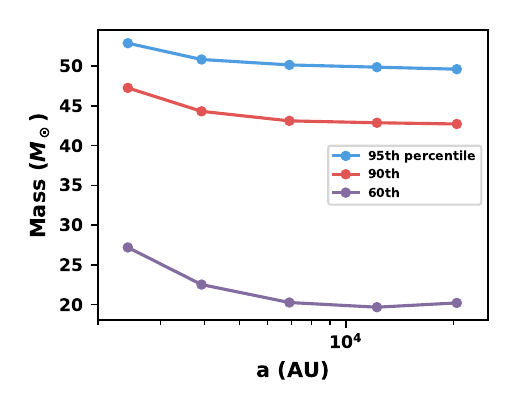}{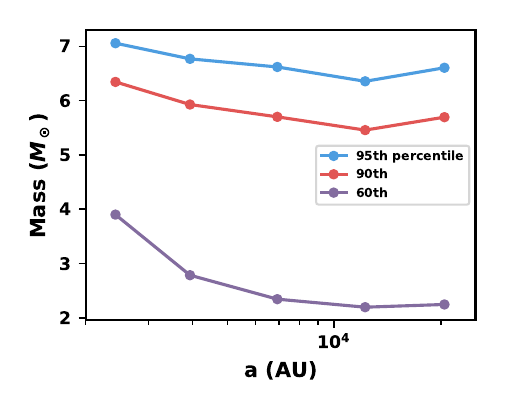}
\caption{Detection limits for companions around a 60$M_{\odot}$ and a 8$M_{\odot}$ star at the LMC distance using the speckle constraints. The lines indicate the 3-$\sigma$, 2-$\sigma$ and 1-$\sigma$ mass detection limits. \label{molusc}}
\end{figure*}

\subsection{Monte Carlo Simulations}\label{sec:molusc}

Speckle contrast curves can be used to estimate the masses of undetected physical companions, for example, by comparing the absolute magnitude of the detection limits with isochrones, and extracting the predicted mass from the isochrone for that luminosity, assuming a main sequence companion (e.g. \citealt{salinas20}). More detailed predictions can be obtained by computing stellar orbits of different companion masses, and comparing them with the observational constraints put by the data. This can be done, for example, via Montecarlo simulations using \textsc{molusc} \citep{wood21}\footnote{https://github.com/woodml/MOLUSC}.

We modeled two such cases, one with a primary mass of 8$M_{\odot}$, and another with 60 $M_{\odot}$, encompassing the mass range expected for these mYSOs \citep{whitney,sewilo13}. For the \textsc{molusc} input orbital parameters we used: a) a log-normal period distribution \citep{raghavan10} restricted to semi-major axes between 2000 AU$<a<25000$ AU, b) a uniform mass ratio distribution, from $q=m_2/m_1$=0.1 to 1, c) an orbital inclination uniform in $\cos(i)$, d) eccentricities from a uniform distribution between 0 and 1,  and e) uniform distribution for both the pericenter phase and the argument of periapsis \citep[see][for details]{wood21}.
For each case, 1 million companions were generated. To establish the relation between masses and magnitudes, a PARSEC \citep{bressan12} isochrone with 1 Myr and [M/H]=--0.5 was used.

Fig.\,\ref{molusc} shows the results for both cases, when taking the contrast curve of LMC J045403.62-671618.2 as an example. The semi-major axes of the orbits that are not ruled out by the contrast curve are binned, and the $95^{th}$, $90^{th}$ and $60^{th}$ percentiles are calculated for the surviving companions in each bin. Our results indicate that between 2000--25\,000\,au equal mass binaries are within the detection limit of ZORRO, and their absence suggests a physical lack of such companions in the sample. Binaries with $q$\,=\,0.5, would be detected at a 60\% confidence level, so binaries below such a mass ratio would most likely be undetected using the current observational techniques for this sample. Based on these simulations, we can rule out the presence of undetected equal mass wide binaries in the sample, and those with a 0.5 mass ratio at the 60\% confidence level.

\section{Discussion} \label{sec:syle}

\subsection{Comparison of the binary fraction as a function of metallicity}

We imaged 34 mYSOs (23 in the LMC; 11 in the SMC) using ZORRO. Our results found three wide binary candidates in the LMC, one of which is a chance alignment. No wide binaries in the SMC were present in our sample within the detection limits. Our analysis also indicates that our YSO sample is contaminated by around 35\%; likely by luminous giants which can be hard to distinguish from mYSOs at these distances even using optical and mid-infrared spectroscopy (e.g. \citealt{kalari14, jones17}). Based on these considerations, we argue that the wide binary fraction of our sample, for companions between $\sim$20--1000\,mas (or physical separations between in the LMC and SMC between $\sim$2000--20000\,au) at contrasts less than 3\,mag are 9$\pm$5\% in the LMC, and $<5$\% in the SMC.

\begin{figure}
\includegraphics[width=1\columnwidth]{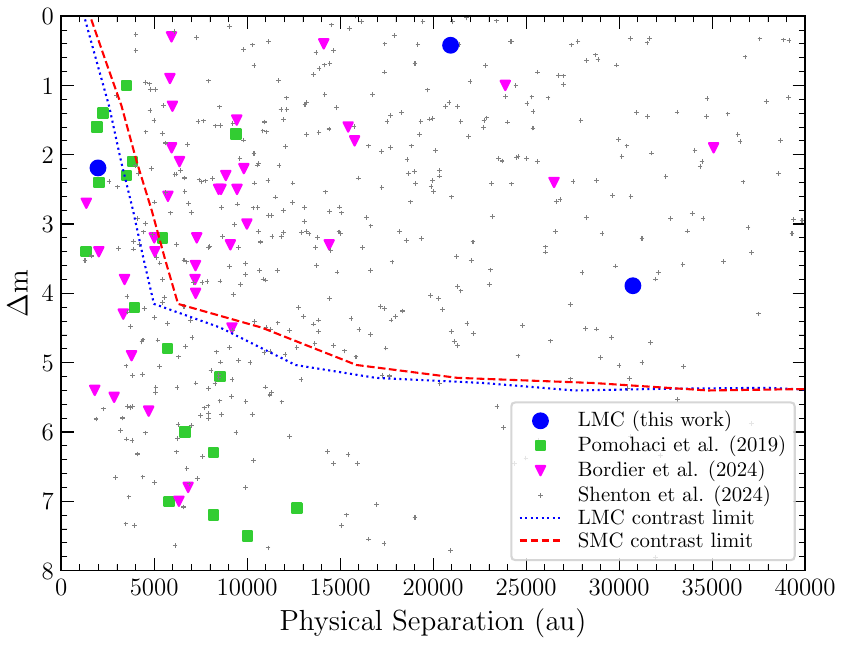}
\caption{Detected binaries in our sample shown by blue circles, compared to the ones detected in Pomohaci et al. (2019), and our observational parameter space. $\Delta m$ refers to difference in the magnitude in the same filter as the primary, but not necessarily the same as comparative studies. All targets shown are only high mass YSOs. \label{money}}
\end{figure}

These results present a stark contrast to those found in the Milky Way for wide binaries. An adaptive optics imaging study in the $L$-band by \cite{bordier} found a high binary fraction of 62\% for 13 known mYSOs. For a fair comparison, we constrain the Galactic results to our physical separation and contrast results as given in Fig.\,\ref{money}. Computing the binary fraction of \cite{bordier} based on the physical space probed by this study (that is within the contrast curves shown in Fig.\,\ref{money}), we find the same binary fraction. This is because the binaries they detect outside our parameter space are higher order multiples with other companions within our parameter space. Similarly, \cite{shenton} attempted to detect wide companions around a sample of 402 mYSOs using near-infrared $K$-band surveys. They found a binary fraction around 62\%, when combining multiple surveys. In the same physical space as our study, their binary fraction becomes 54\%. This suggests a significant fraction of binaries detected in both the near and mid-infrared in the Galaxy should have been detected using ZORRO given the physical separation as a limit, and assuming that the difference in magnitude across the optical and infrared of the companions is similar. The study of \cite{pomohaci} found a lower binary fraction from a sample of 32 mYSOs of 32\% using adaptive optics $K$-band imaging. Applying our physical constraints on their sample, their binary fraction decreases to 13\%, which is in keeping with the binary fraction observed here. The comparisons are summarized in Fig.\,\ref{money2}. 

To estimate the statistical significance of our result we utilize bootstrapping techniques. For the binary fraction estimated by each study, we compute the distribution of binary fractions assuming 100000 iterations but only a sample size of 23 (corresponding to our LMC sample size). The resulting probability distribution, and the 25\% and 75\% quantiles are shown in Fig.\,\ref{money2} along with the estimated binary fraction. These values are then used to compare the probability that the two underlying distributions are drawn from the same sample using a Kolmogorov-Smirnov two sample test. Comparing the LMC and SMC distribution with those in the Galaxy, the $p$-value is less than 0.05 for all Galactic studies suggesting we can reject the hypothesis that the underlying distributions are the same. 

From the above considerations, we suggest that when analyzing the same physical separations as Galactic studies, we find a dearth of wide companions in the Magellanic Clouds. In this section, we explore the possible explanations for the low wide binary fraction.

\subsubsection{Selection effects and biases} \label{sec:stfle}
Our sample primarily consists of the brightest optical targets from an infrared-selected sample. This would naturally push us towards the most massive, and most evolved of our targets. It would also preferentially select binaries. Therefore it is surprising to not observe a higher fraction of binaries in our sample. 

It is probable that the data cannot sufficiently resolve the outer disc and binary configurations for all cases. Since speckle imaging detects companions using the power spectrum, dealing with non-point sources in the final image analysis is not straightforward. However, the pipeline is able to identify extended structures and asymmetries as seen in \cite{shara}. In fact, the outer disc radius of most YSOs is a unlikely to be larger than a 1000\,au, beyond which  asymmetries due to discs (at a variety of inclination angles) may have been detected. Most importantly, any binary formed must have cleared much of the disc or envelope material around its immediate vicinity given the more evolved nature of our sample. Therefore, it would be unlikely that a significant proportion of binaries are not detected due to dusty discs, or envelopes. 

A significant effect is caused from small number statistics. There are 742 high probability mYSOs in the SMC in the sample of \cite{sewilo13}, and 457 high probability mYSOs in the LMC \citep{whitney}. When sorting by mass, there are at least 168 mYSOs in the whole SMC based on the mass estimates of \cite{sewilo13}. Therefore, our sample is extremely small ($\lesssim$5\% in both galaxies), and may not effectively translate to large number statistics. For a 3$\sigma$ confidence on our computed binary fraction, we effectively have a $\sim$35\% error bar. Therefore, to endorse our results with statistical significance we need to increase in an unbiased fashion  our sample size  by an order of magnitude in both galaxies.



\subsubsection{Physical effects} \label{sec:styl}

Here we only consider the physical effects specific to individual sources, such as the extinction or environment, rather than global properties such as metallicity.

Firstly, dense environment dynamics may lead to a lower binary fraction. For e.g., the close binary fraction in the 30\,Doradus region is smaller by about 20 percent than the rest of LMC, as found by \cite{sana13, dunstall15} even at young ages. It is very likely the such binary fractions translate to wide binaries. However, a less than 30\% of our sample is located in dense environments (where $\rho>$2000 sources\,arcmin$^{-2}$). 

Secondly, extinction may cause some bias in our sample. Wide binaries in our sample may not be detected in a filter/field because they are beyond our detection limit due to extinction. However, this would imply extinction differences of greater than a few mag within scales less than 1$\arcsec$ (0.25\,pc). While such scales are seen in mYSOs that are very young and dense environments, this is not the majority of our sample. This is also beyond the scales which the disc, or envelope provides significant extinction. Therefore, while the local environment or extinction may affect a few specific cases, it is unlikely that they explain the lower binary fraction of our sample. 


\begin{figure}
\includegraphics[width=1\columnwidth]{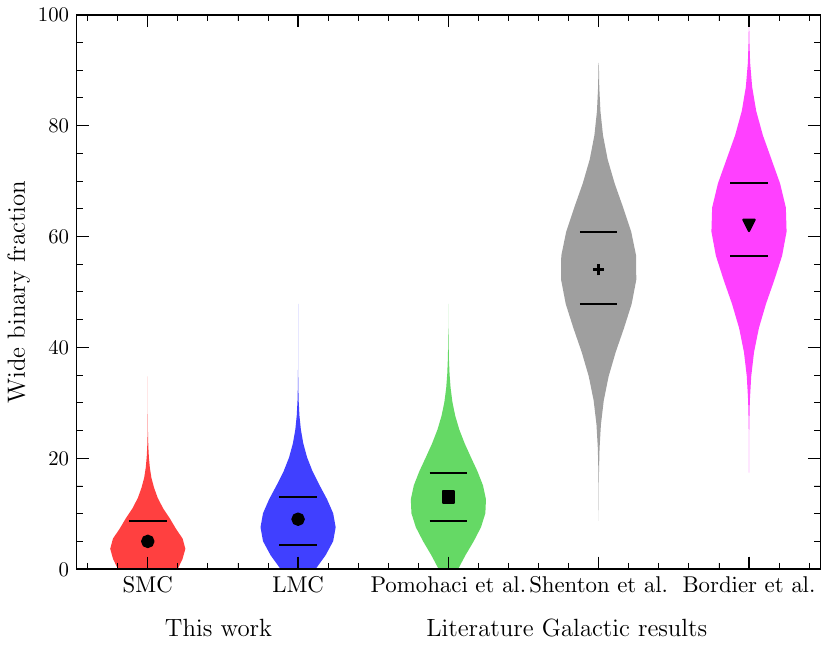}
\caption{Violin plot of wide binary fraction of mYSOs in the LMC and the SMC limit. The wide binary fraction from multiple Milky Way studies, but in the physical parameter range of ZORRO at the distance of the LMC is also shown for comparison. Colors are same as Fig.\,4 with the SMC shown in red. The solid lines giving the 25\% and 75\% quantile ranges. The bootstrapped probability distribution computed if the observed sample size is equal to the LMC sample size but with the underlying binary fraction reported by each study is shown by the bodies. The scatter points give the binary fraction of each distribution. \label{money2}}
\end{figure}

\subsubsection{The low wide binary fraction is real} \label{sec:style}

We suggest that although physical or selection effects may affect the binary fraction observed negatively, this does not apply to the majority of our sample. A significant fraction of mYSOs in our sample do not have wide binaries. This suggests an absence of wide binaries in these low-metallicity environments from preliminary results, and may need theoretical explanations (as seen in Fig.\,\ref{money2}). 

Theoretically, metallicity increases the cooling of the collapsing protostellar gas and associated protostellar disks. There is less molecular and dust cooling in more metal-poor gas, such that we would expect wider separations of visual binaries in more metal-poor systems. The outer disks around massive stars have to be cold enough (10-30K) to fragment by gravitational instability, and it is possible that very metal-poor protostellar disks (like in the SMC) may be too warm to fragment at all. If protostellar disks do fragment, they would do so at larger distances from the heat source. There exist detailed simulations of the disc temperatures and radii as a function of metallicity to understand fragmentation (such as in \citealt{sugimura}), which suggest that exteremly metal poor massive stars form wide binaries in the separation ranges between 2000--20000\,au. In addition, at lower metallicities, magnetic braking is weaker and carries away lesser angular momentum, which may lead to wider and more massive binaries with increased rotation rates \citep{zin03}. Therefore, theoretically, one would expect a higher wide binary fraction at lower metallicities. It is interesting to note that recent interferometric observations of close binaries, in the 2-300\,au range predict a much smaller binary fraction in the Milky Way, around 17--25\% \citep{koumpia}. 

Wide binaries around mYSOs are not detected at the same frequencies in the Magellanic Clouds, compared to the Milky Way. It is possible that the binaries are formed on smaller scales than can be resolved by the data, or that the sample size is insufficient. But, if these results are seen in a larger statistical sample of mYSOs in the Magellanic Clouds, there will exist a need to radically revisit our idea of how metallicity affects the star formation process, especially that of wide binaries. The implications on the binary fraction of metal poor stars is important. For e.g., metal-poor binaries in the early Universe may have evolved into binary black holes which are now detectable by gravitational waves, or the first X-ray binaries formed provide a crucial heating source to the intergalactic medium. Such events can inject significant matter and energy into the Universe and affect both stellar and galactic evolution.

An observational estimate on the wide bianry fraction at lower metallicities is also important to inform simulations considering the formation of the first stars \citep{sugimura}. Therefore, to validate our suppositions, we suggest a much larger and comprehensive follow-up to estimate the wide binary fraction of mYSOs, both in the Magellanic Clouds and the Milky Way. 



\section{Conclusions}

This study presents observational results indicating a significantly lower wide binary fraction in massive young stellar objects in the Magellanic Clouds compared to the Milky Way. This suggests that metallicity may play a crucial role in shaping the formation of wide binaries, and is contrary to current theoretical expectations. Further follow-up is essential to improve our understanding of binary formation in such environments and provide insights into the global implications.


\begin{acknowledgments}

V.M.K. thanks Lila and Asha Kalari for help with the figures, and Imogen Kalari for proof reading. We thank the referee for a positive and constructive report. Observations in the paper made use of the High-Resolution Imaging instrument Zorro. Zorro was funded by the NASA Exoplanet Exploration Program and built at the NASA Ames Research Center by Steve B. Howell, Nic Scott, Elliott P. Horch, and Emmett Quigley. Zorro was mounted on the Gemini South telescope of the international Gemini Observatory, a program of NSF’s NOIRLab, which is managed by the Association of Universities for Research in Astronomy (AURA) under a cooperative agreement with the National Science Foundation on behalf of the Gemini Observatory partnership: the National Science Foundation (United States), National Research Council (Canada), Agencia Nacional de Investigaci\'{o}n y Desarrollo (Chile), Ministerio de Ciencia, Tecnolog\'{i}a e Innovaci\'{o}n (Argentina), Minist\'{e}rio da Ci\^{e}ncia, Tecnologia, Inova\c{c}\~{o}es e Comunica\c{c}\~{o}es (Brazil), and Korea Astronomy and Space Science Institute (Republic of Korea).

\end{acknowledgments}

%

\vspace{5mm}





\clearpage

\appendix

\section{Observed targets}

\begin{deluxetable}{lllll}
\tablenum{1}
\tablecaption{Catalog of observed sources}
\tablewidth{0pt}
\tablehead{
\colhead{Adopted Name{$^1$}} &
\colhead{Right Ascension} &
\colhead{Declination} &
\colhead{$V$} &
\colhead{$I$} \\
\colhead{} &
\colhead{(J2000)} &
\colhead{(J2000)} &
\colhead{(mag)} &
\colhead{(mag)} 
}
\startdata
  SMC\_Y37 & 00:36:43.960 & $-$72:37:22.20 & 15.207 & 14.369\\
  SMC\_Y97 & 00:44:32.020 & $-$74:40:29.40 & 15.593 & 14.229\\
  SMC\_Y115 & 00:45:21.310 & $-$73:12:18.60 & 16.464 & 15.604\\
  SMC\_Y120 & 00:45:24.460 & $-$73:12:38.80 & 15.239 & 14.774\\
  SMC\_Y451 & 00:49:27.290 & $-$72:47:38.80 & 15.732 & 15.471\\
  SMC\_Y259 & 00:55:29.570 & $-$71:53:12.20 & 15.271 & 14.295\\
  SMC\_Y635 & 01:03:04.530 & $-$71:31:48.10 & 16.156 & 15.927\\
  SMC\_Y786 & 01:12:22.900 & $-$71:58:20.50 & 16.584 & 15.425\\
  SMC\_Y825 & 01:14:17.260 & $-$73:15:49.40 & 15.049 & 15.039\\
  SMC\_Y880 & 01:20:44.280 & $-$71:47:45.00 & 15.658 & 15.478\\
  SMC\_Y905 & 01:25:17.050 & $-$73:23:06.10 & 14.238 & 14.451\\
  LMC\_J045403.62-671618.2 & 04:54:03.620 & $-$67:16:18.30 & 15.8 & 15.349\\
  LMC\_J045623.22-692748.8 & 04:56:23.230 & $-$69:27:48.90 & 14.628 & 13.123\\
  LMC\_J050338.91-690158.5 & 05:03:38.910 & $-$69:01:58.60 & 15.091 & 14.981\\
  LMC\_J050451.68-663807.5 & 05:04:51.690 & $-$66:38:07.50 & 15.295 & 14.3\\
  LMC\_J050718.32-690742.8 & 05:07:18.320 & $-$69:07:42.90 & 13.566 & 13.457\\
  LMC\_J050830.53-692237.3 & 05:08:30.530 & $-$69:22:37.40 & 14.282 & 13.58\\
  LMC\_J050957.28-692033.2 & 05:09:57.290 & $-$69:20:33.30 & 17.021 & 14.783\\
  LMC\_J051525.46-661904.8 & 05:15:25.470 & $-$66:19:04.90 & 16.126 & 15.868\\
  LMC\_J051624.89-690000.8 & 05:16:24.900 & $-$69:00:00.90 & 14.402 & 14.066\\
  LMC\_J051728.43-664306.8 & 05:17:28.440 & $-$66:43:06.90 & 15.409 & 14.646\\
  LMC\_J051917.30-693147.3 & 05:19:17.310 & $-$69:31:47.30 & 15.536 & 13.996\\
  LMC\_J052002.04-680420.7 & 05:20:02.050 & $-$68:04:20.80 & 16.524 & 15.912\\
  LMC\_J052020.61-693115.3 & 05:20:20.610 & $-$69:31:15.40 & 16.29 & 15.924\\
  LMC\_J052147.05-675656.6 & 05:21:47.050 & $-$67:56:56.60 & 14.362 & 14.597\\
  LMC\_J052220.96-692001.6 & 05:22:20.970 & $-$69:20:01.60 & 15.826 & 14.346\\
  LMC\_J052352.26-694126.9 & 05:23:52.270 & $-$69:41:26.90 & 12.609 & 12.165\\
  LMC\_J053010.87-695038.3 & 05:30:10.870 & $-$69:50:38.40 & 16.568 & 14.483\\
  LMC\_J053206.06-663023.0 & 05:32:06.070 & $-$66:30:23.10 & 16.636 & 15.938\\
  LMC\_J053455.91-694428.7 & 05:34:55.920 & $-$69:44:28.70 & 16.216 & 14.545\\
  LMC\_J053751.00-690933.8 & 05:37:51.010 & $-$69:09:33.90 & 12.946 & 12.464\\
  LMC\_J054042.18-695743.0 & 05:40:42.180 & $-$69:57:43.00 & 14.08 & 13.793\\
  LMC\_J054059.79-693840.1 & 05:40:59.790 & $-$69:38:40.10 & 12.242 & 12.308\\
  LMC\_J054341.03-710511.2 & 05:43:41.040 & $-$71:05:11.20 & 16.643 & 13.911\\
\enddata
\tablecomments{{$^1$} Designated name is adopted from \cite{whitney} for LMC targets, and \cite{sewilo13} for SMC targets.}
\end{deluxetable}

\section{Reconstructed images and contrast curves}

\begin{figure}
\includegraphics[scale=0.4]{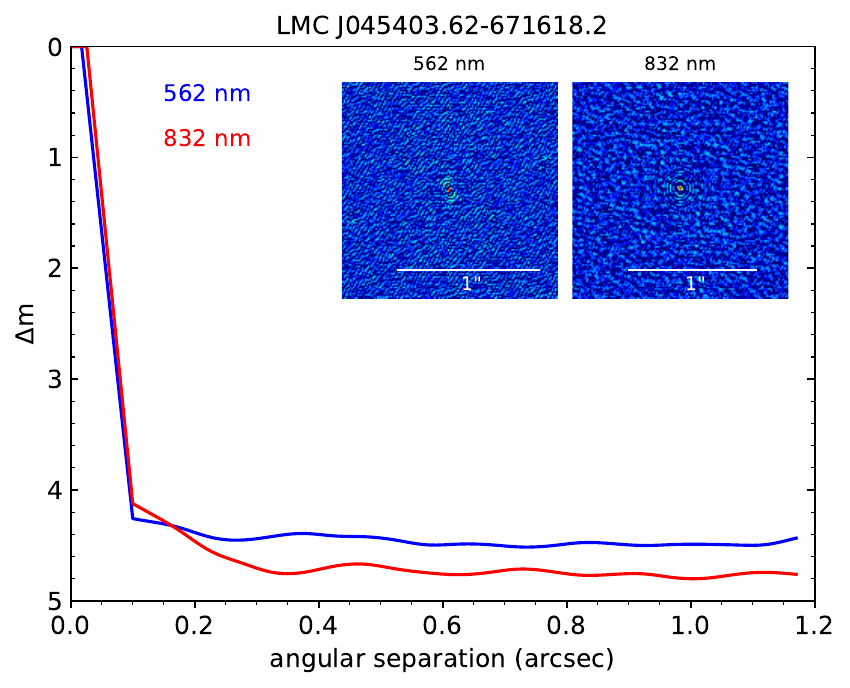}
\includegraphics[scale=0.4]{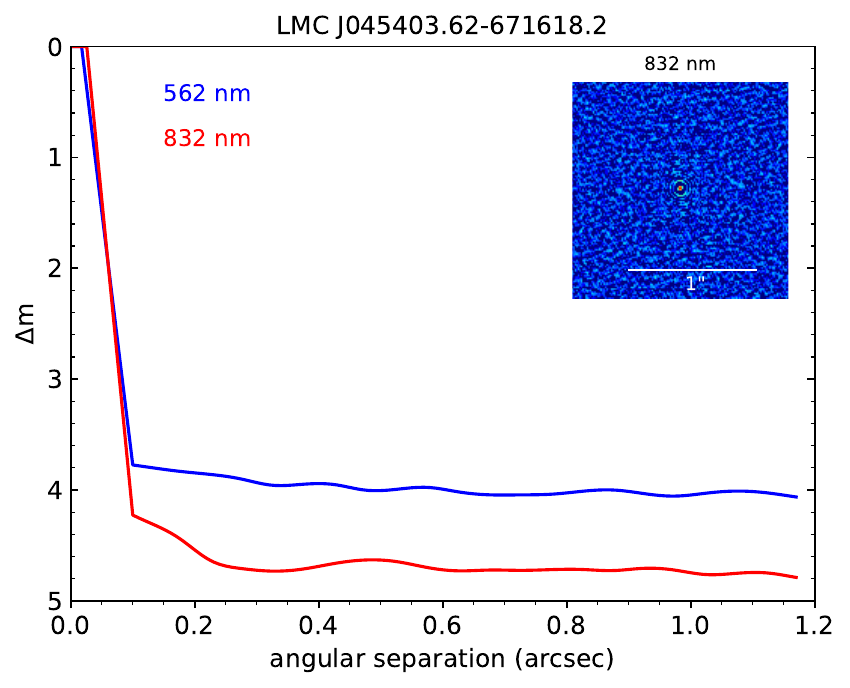}
\includegraphics[scale=0.4]{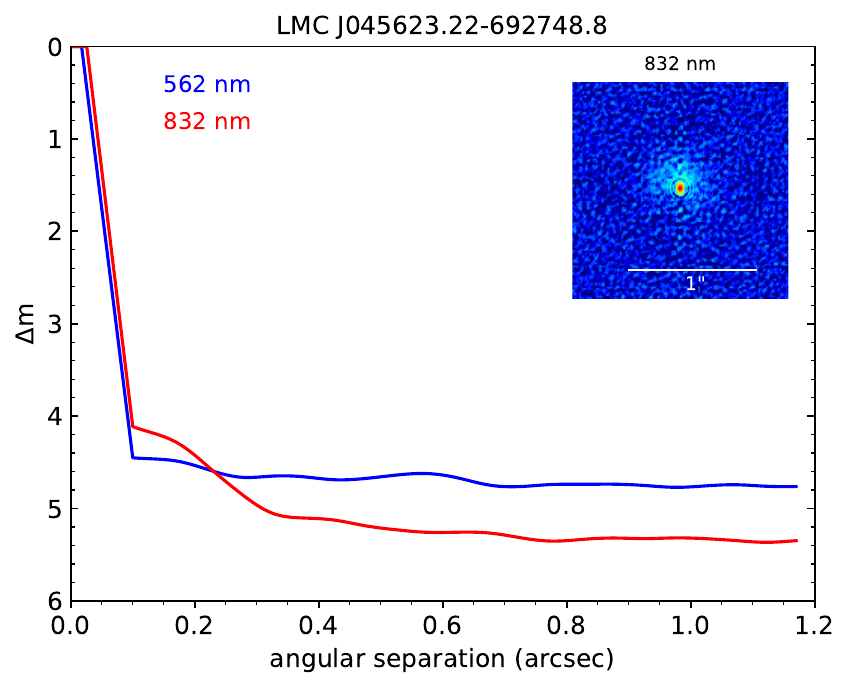}
\includegraphics[scale=0.4]{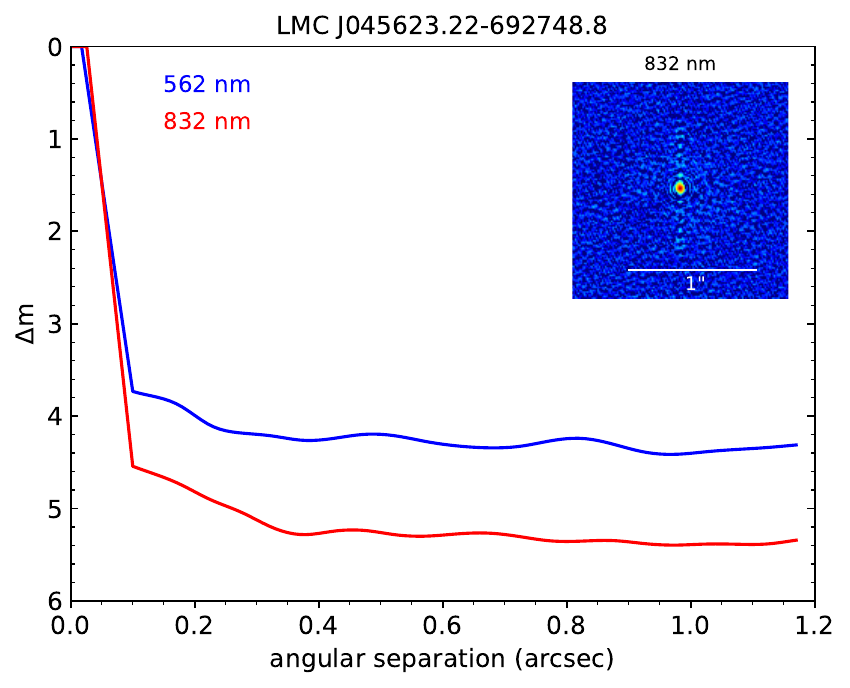}
\includegraphics[scale=0.4]{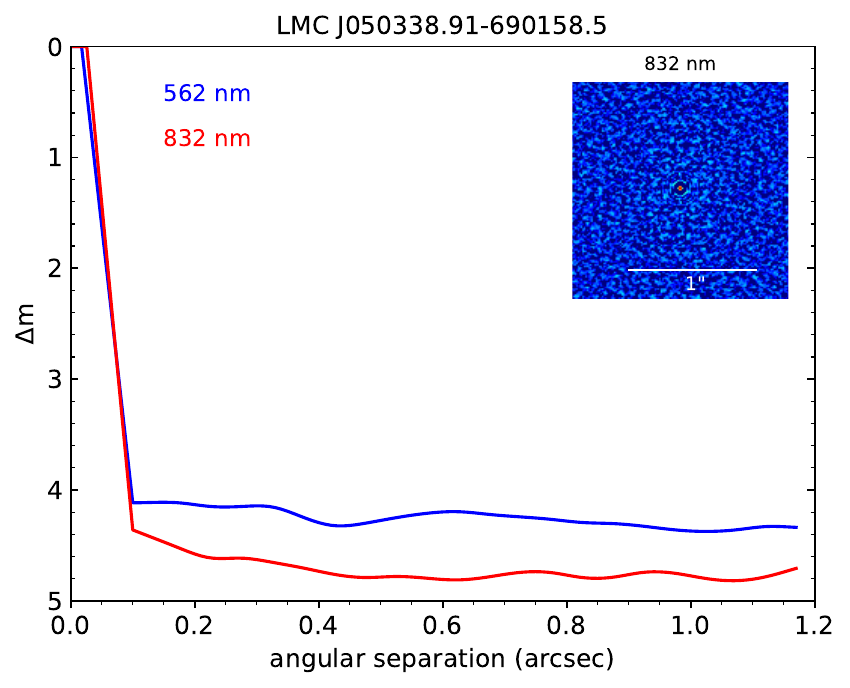}
\includegraphics[scale=0.4]{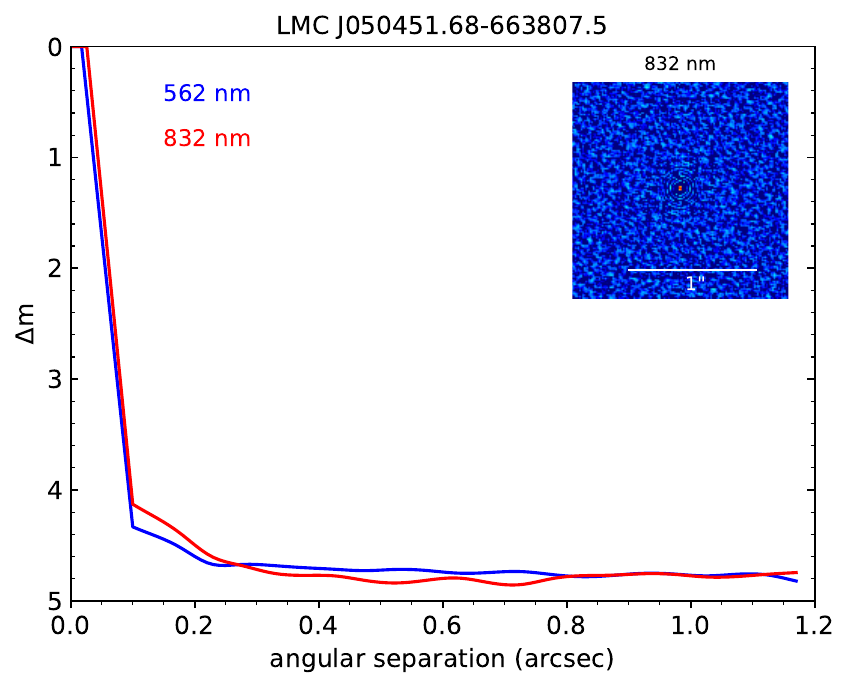}
\includegraphics[scale=0.4]{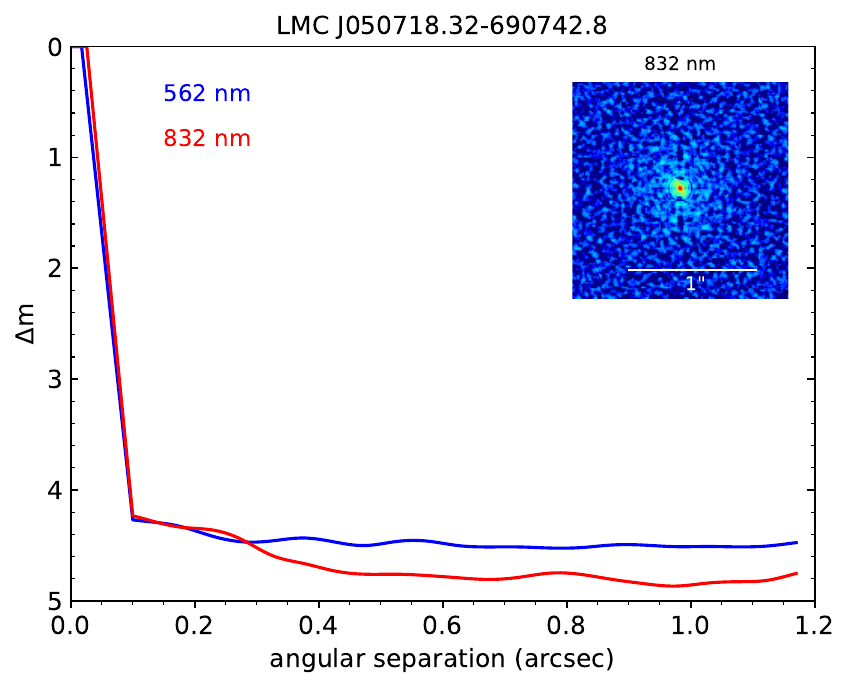}
\includegraphics[scale=0.4]{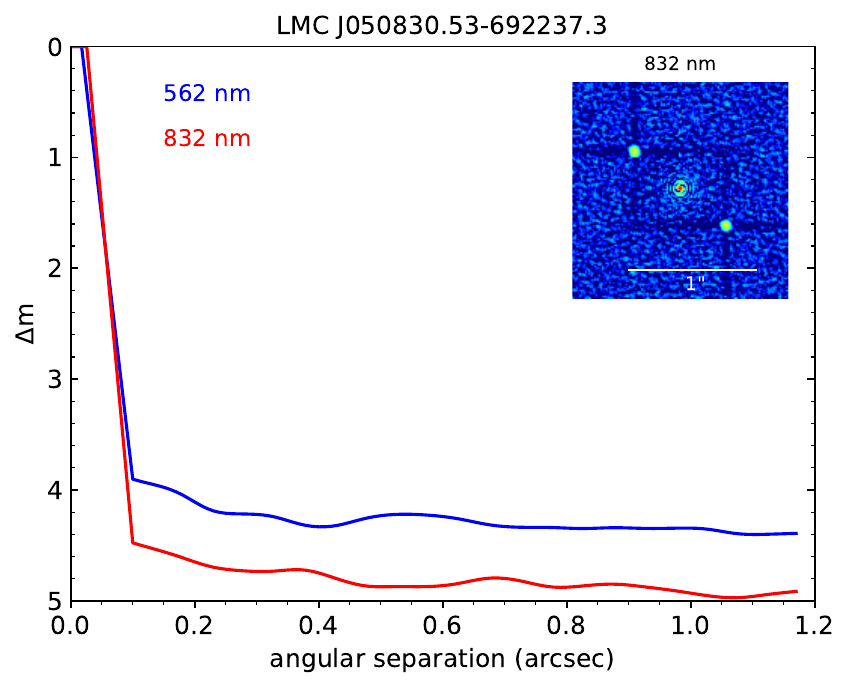}
\includegraphics[scale=0.4]{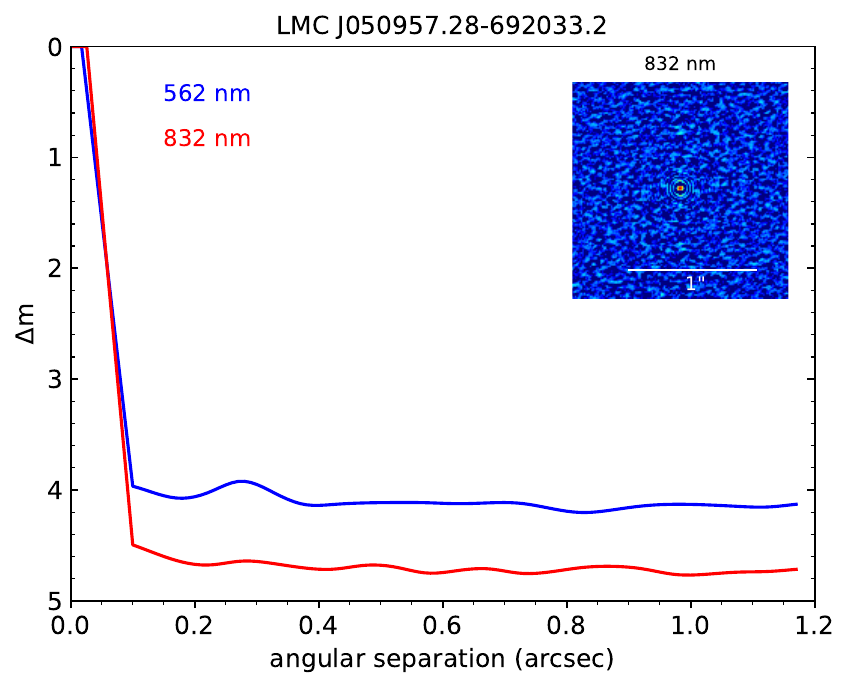}
\includegraphics[scale=0.4]{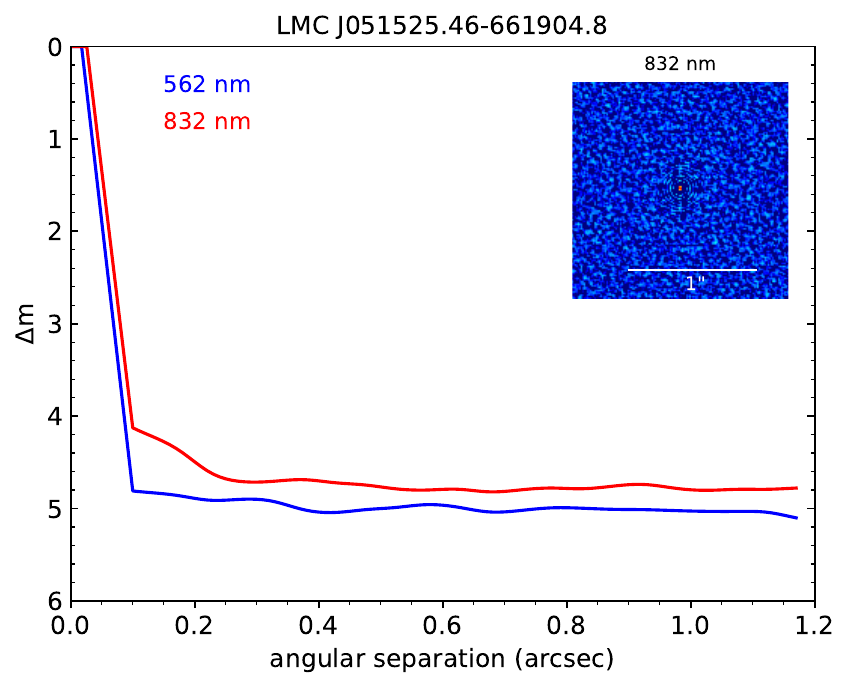}
\includegraphics[scale=0.4]{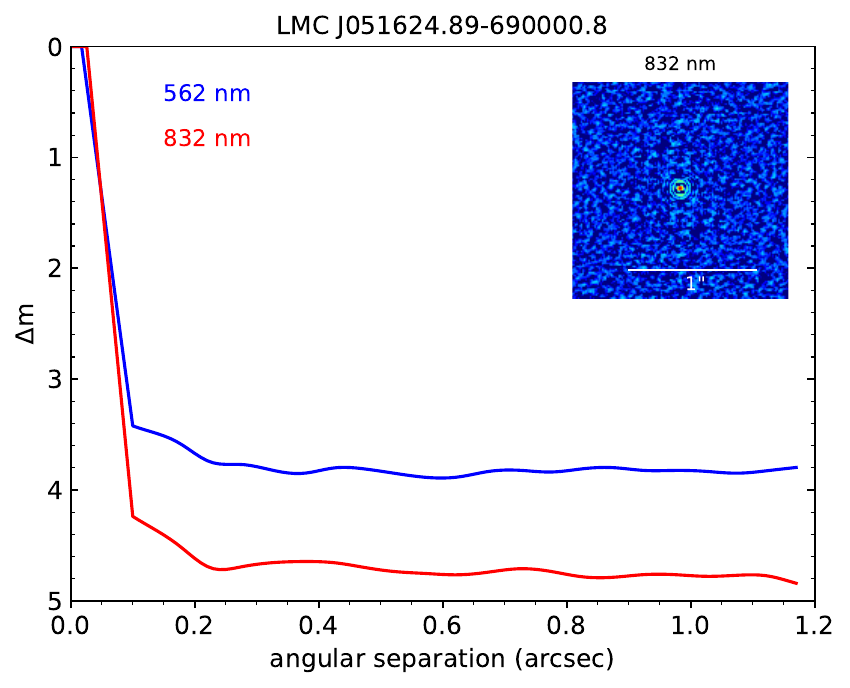}
\includegraphics[scale=0.4]{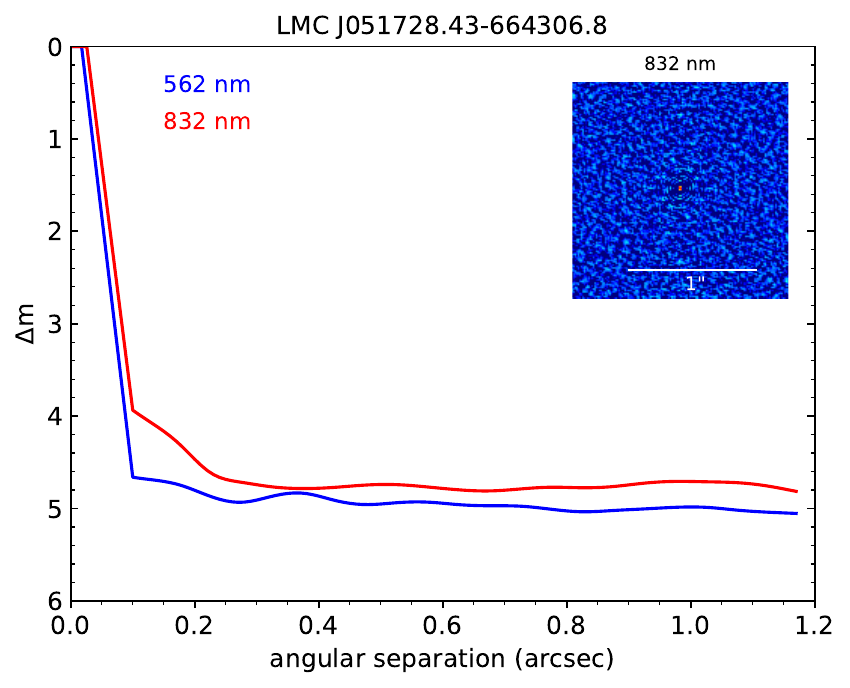}
\includegraphics[scale=0.4]{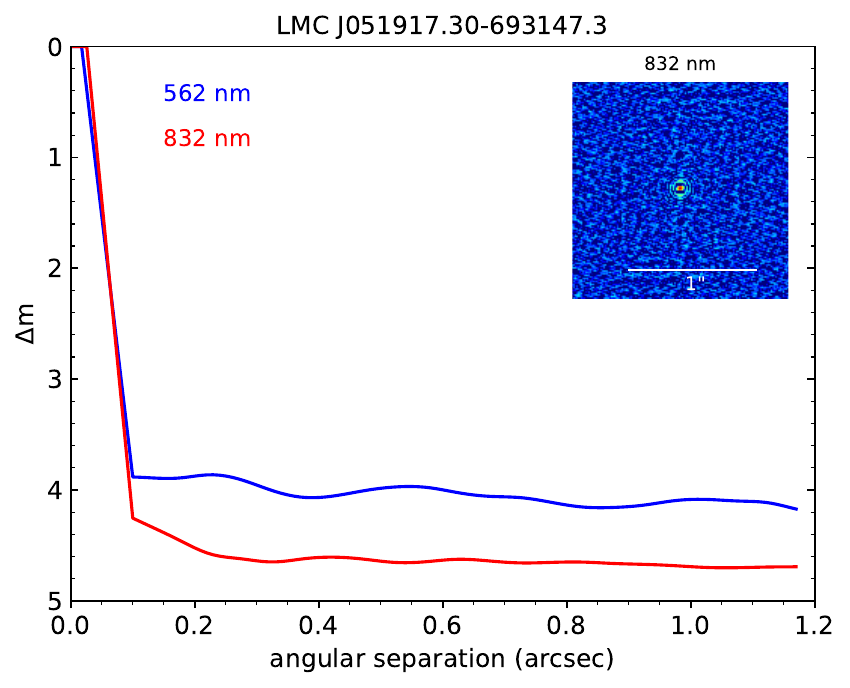}
\includegraphics[scale=0.4]{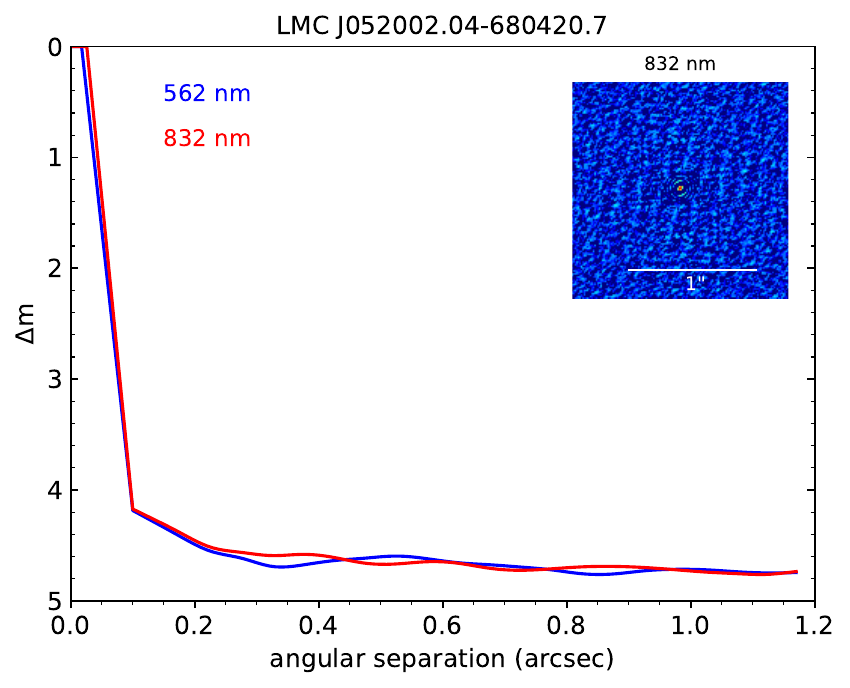}
\includegraphics[scale=0.4]{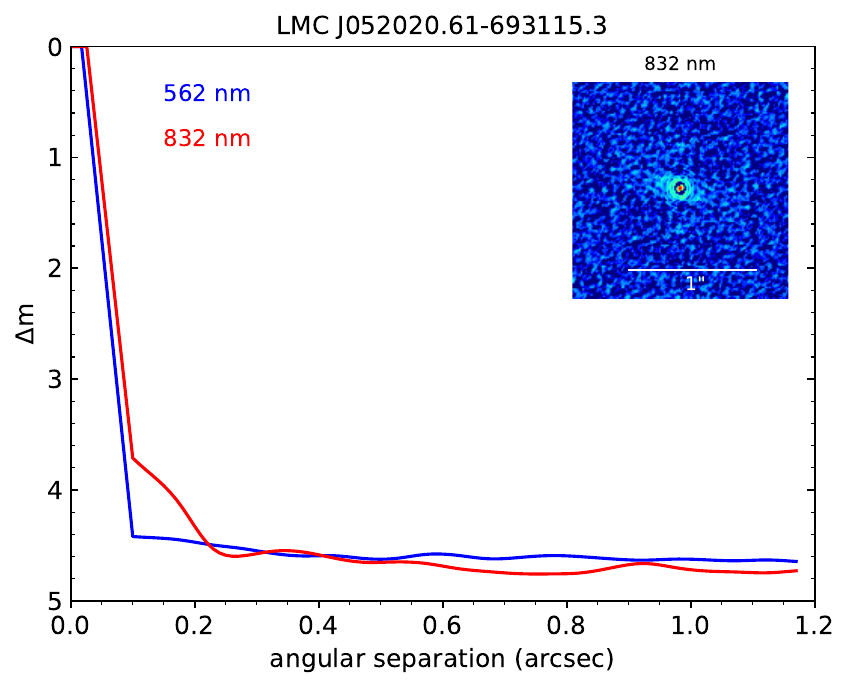}
\end{figure}
\begin{figure}
\includegraphics[scale=0.4]{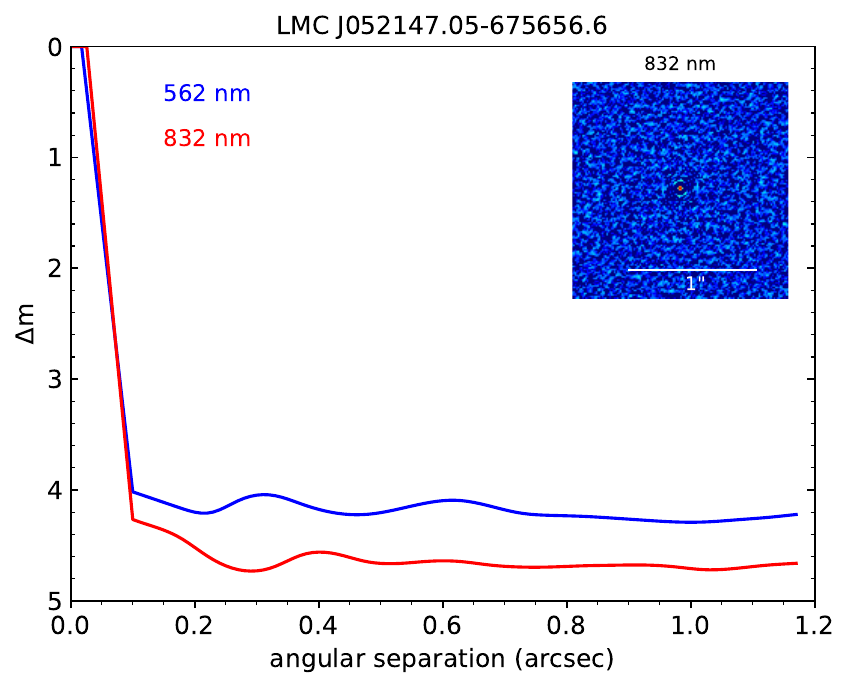}
\includegraphics[scale=0.4]{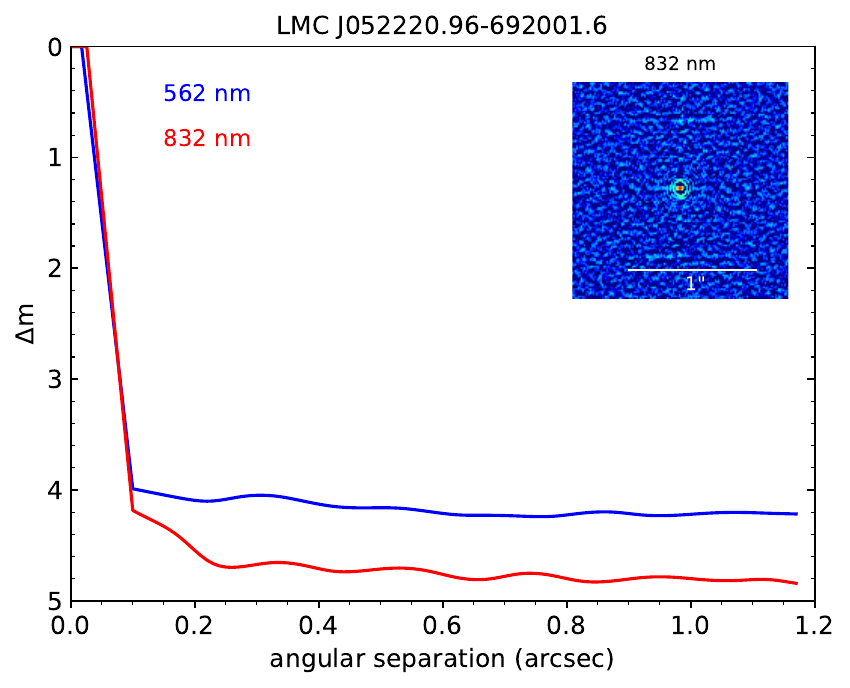}
\includegraphics[scale=0.4]{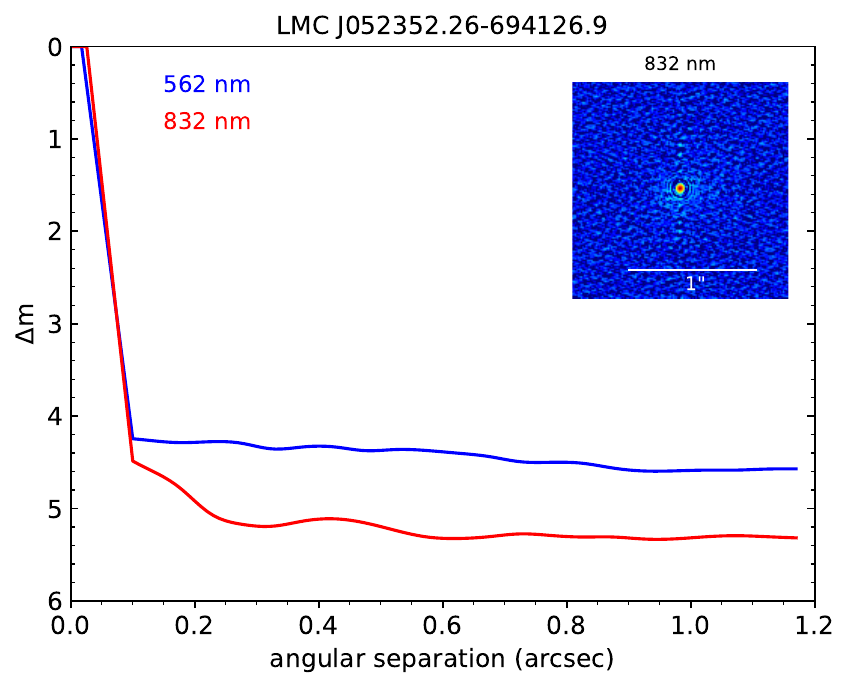}
\includegraphics[scale=0.4]{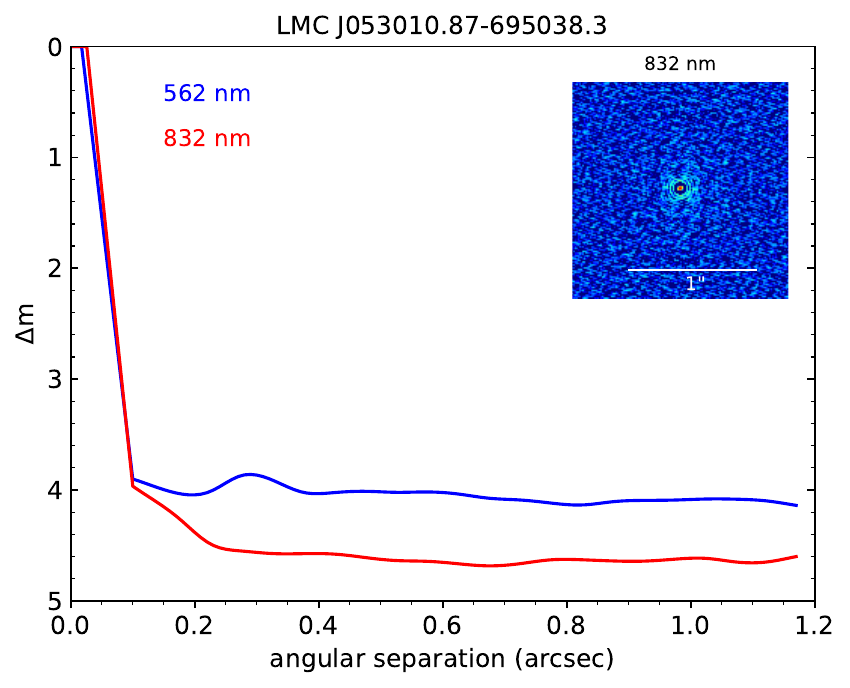}
\includegraphics[scale=0.4]{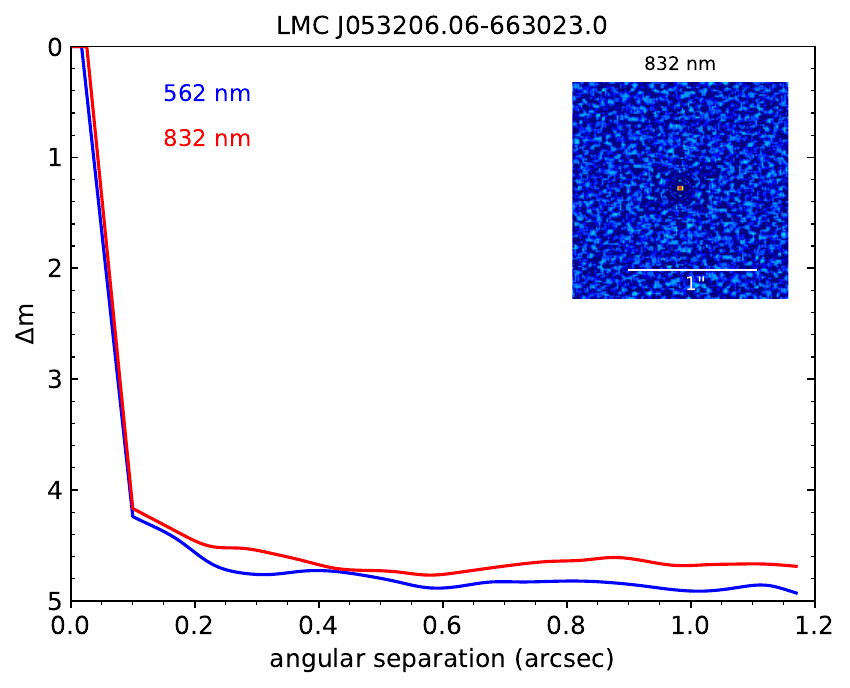}
\includegraphics[scale=0.4]{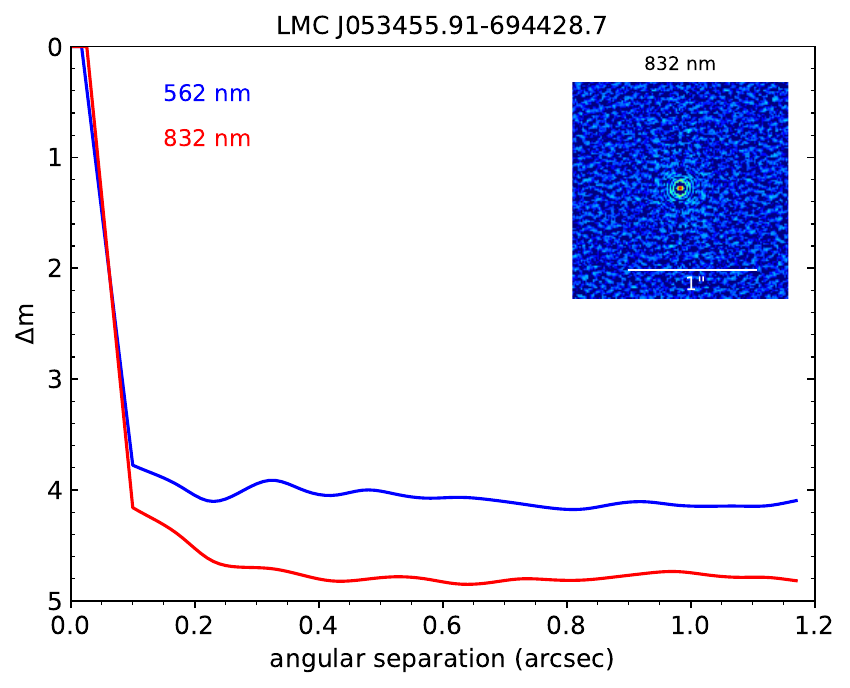}
\includegraphics[scale=0.4]{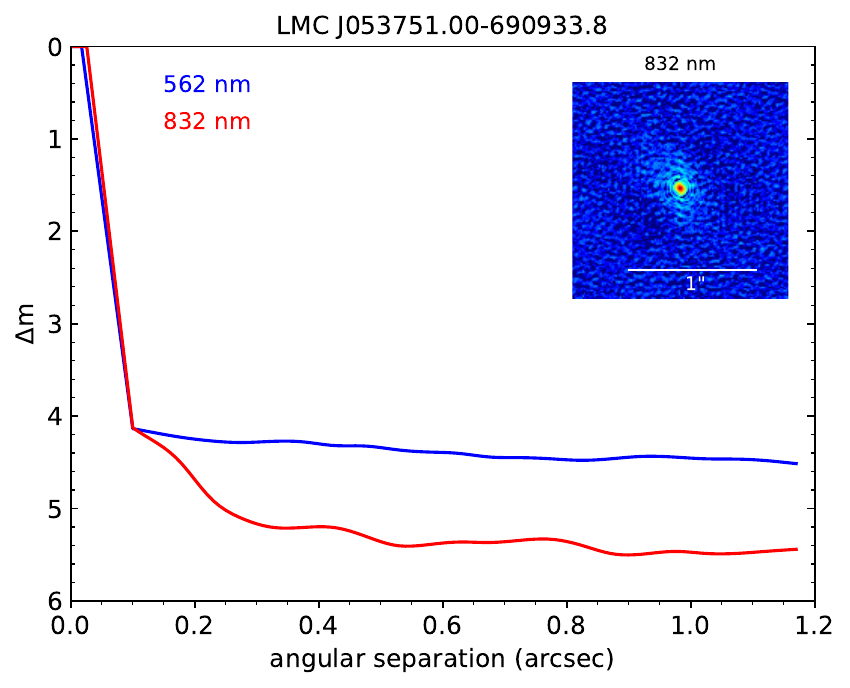}
\includegraphics[scale=0.4]{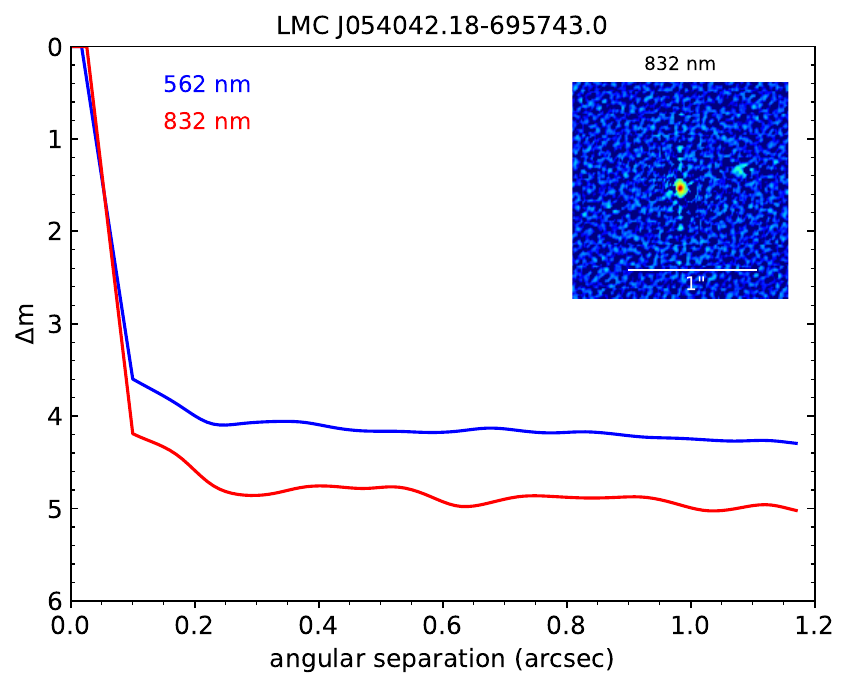}
\includegraphics[scale=0.4]{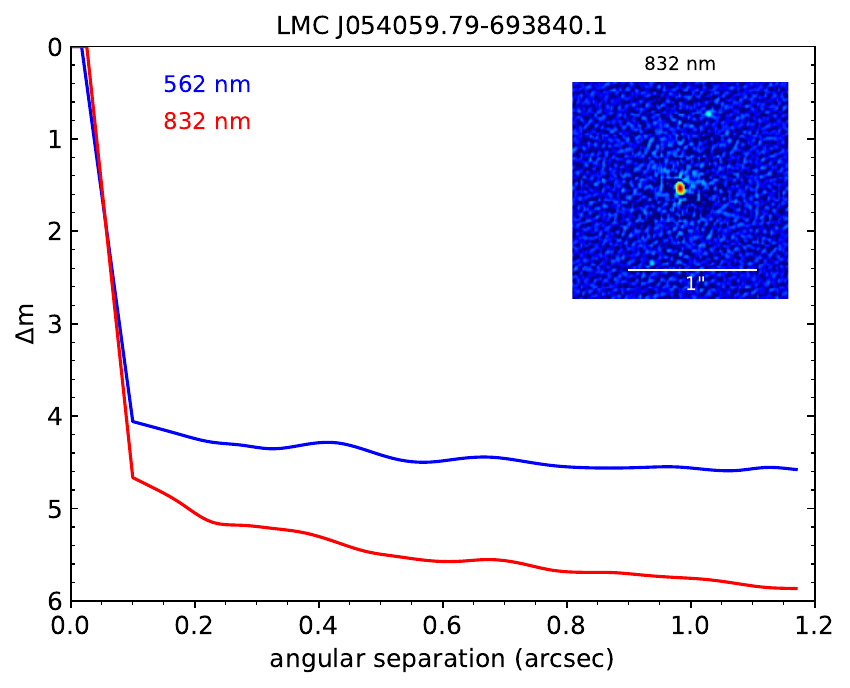}
\includegraphics[scale=0.4]{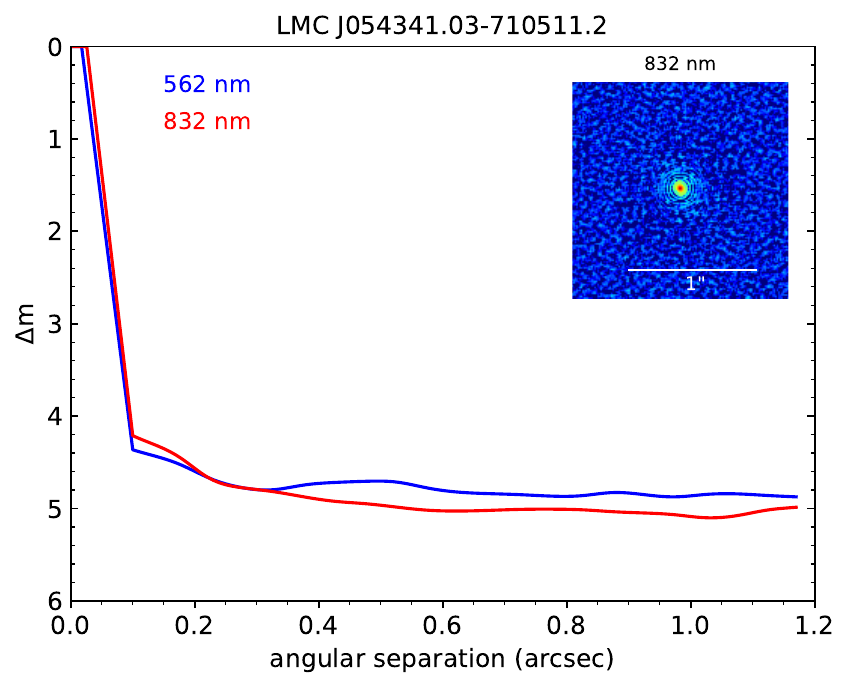}
\includegraphics[scale=0.4]{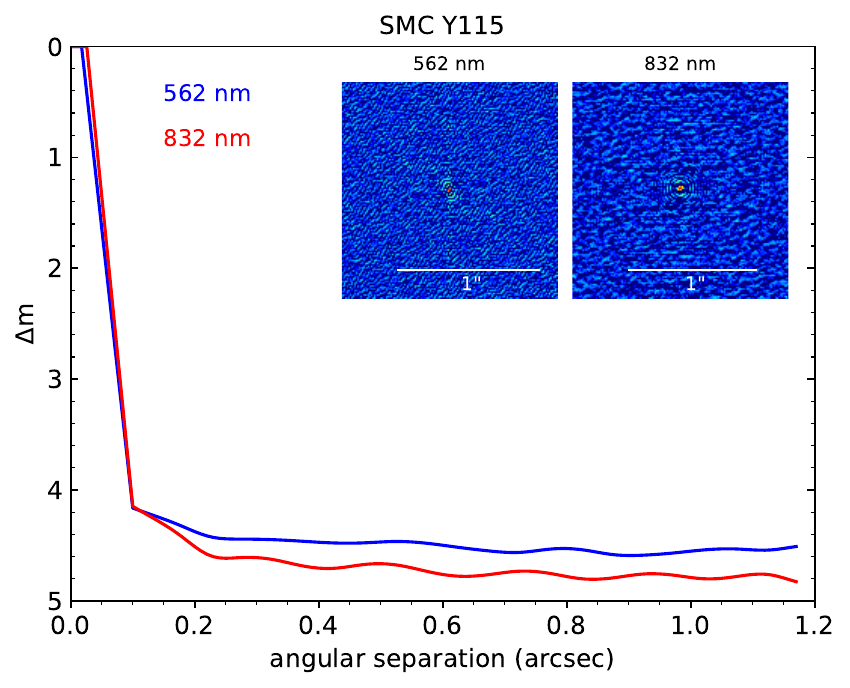}
\includegraphics[scale=0.4]{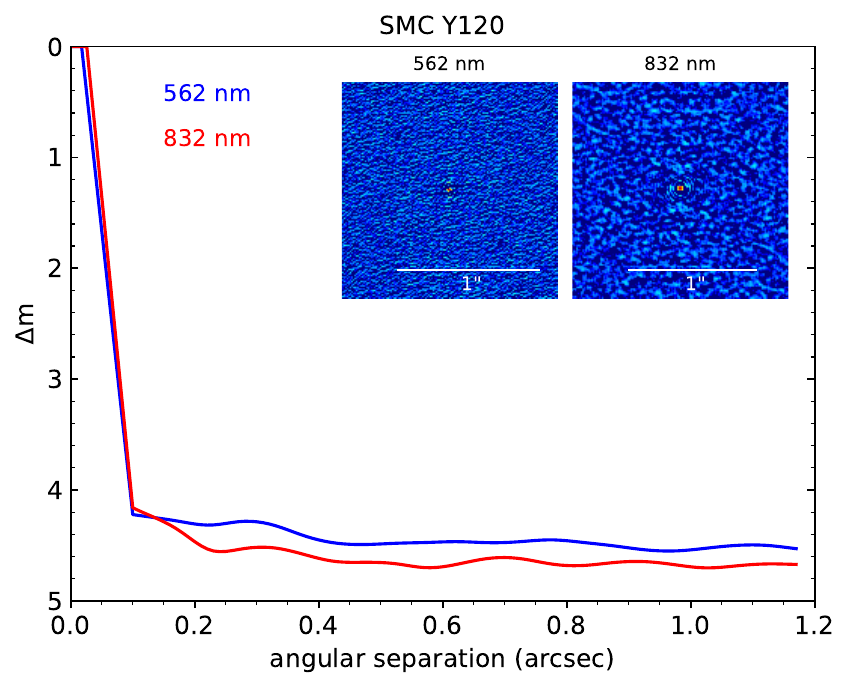}
\includegraphics[scale=0.4]{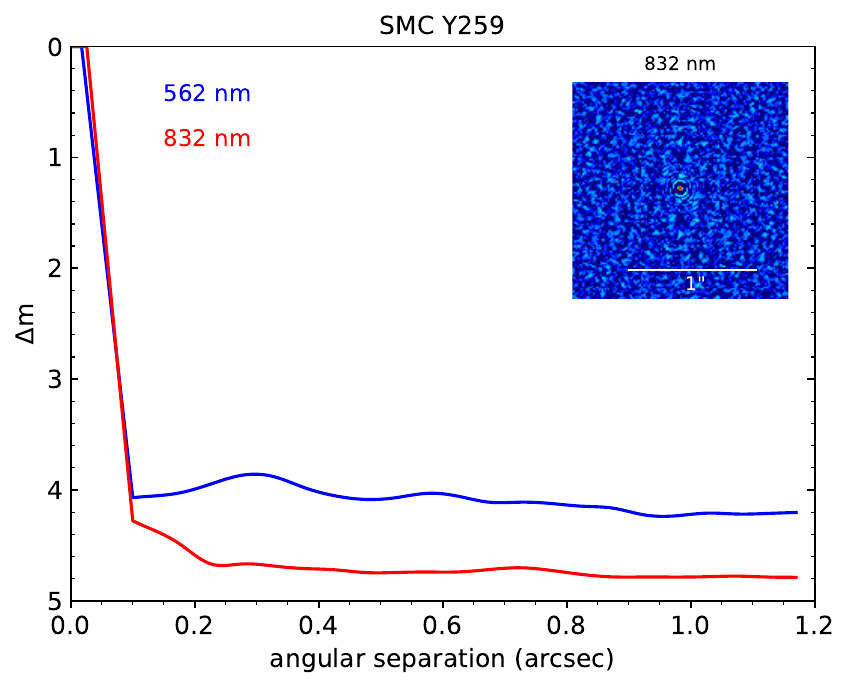}
\includegraphics[scale=0.4]{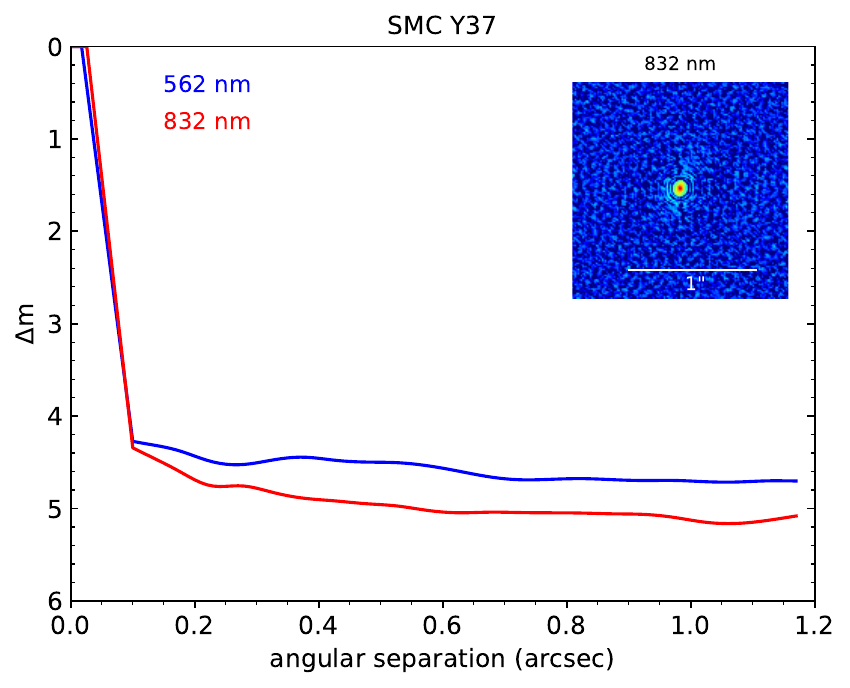}
\includegraphics[scale=0.4]{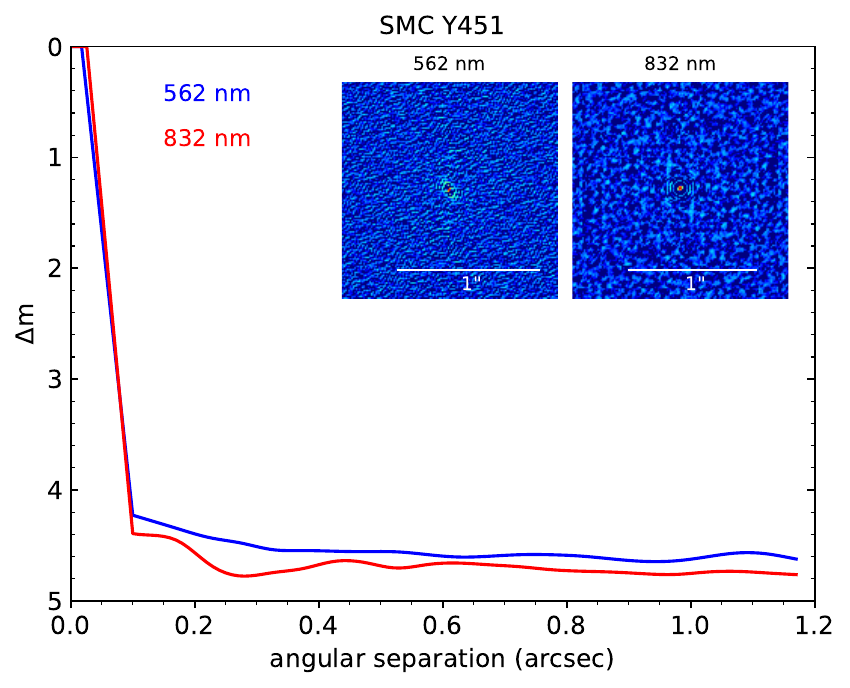}
\end{figure}
\begin{figure}
\includegraphics[scale=0.4]{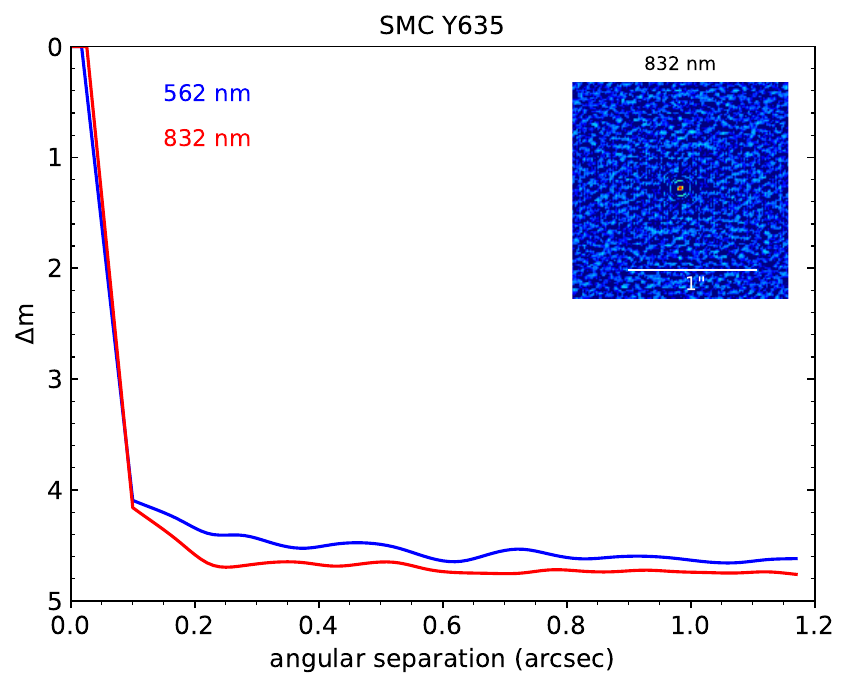}
\includegraphics[scale=0.4]{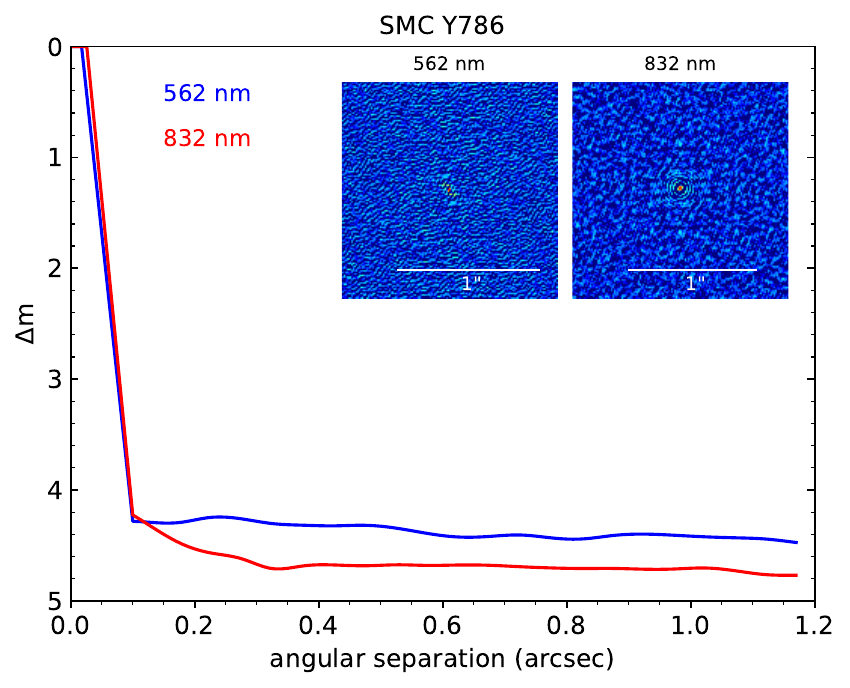}
\includegraphics[scale=0.4]{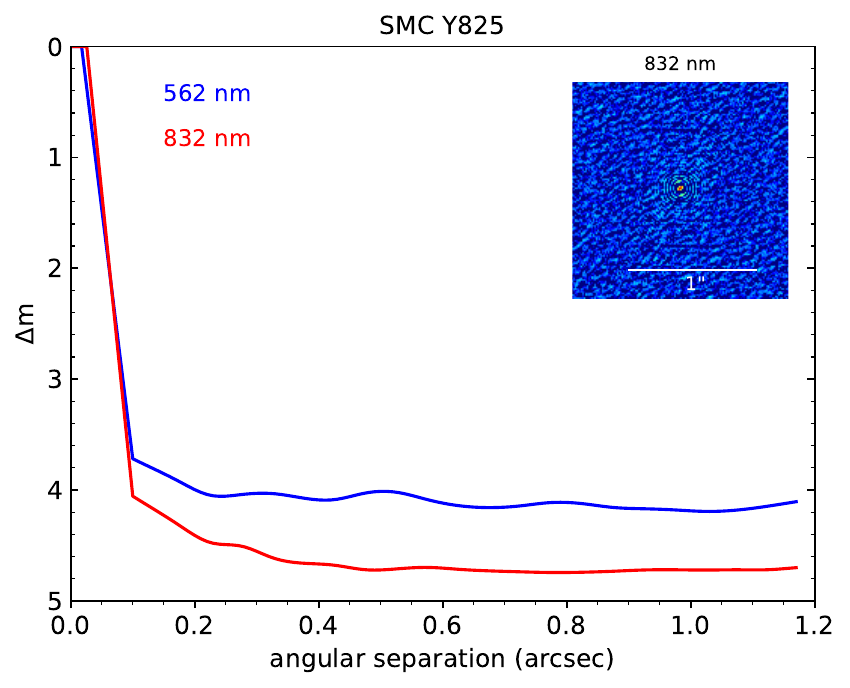}
\includegraphics[scale=0.4]{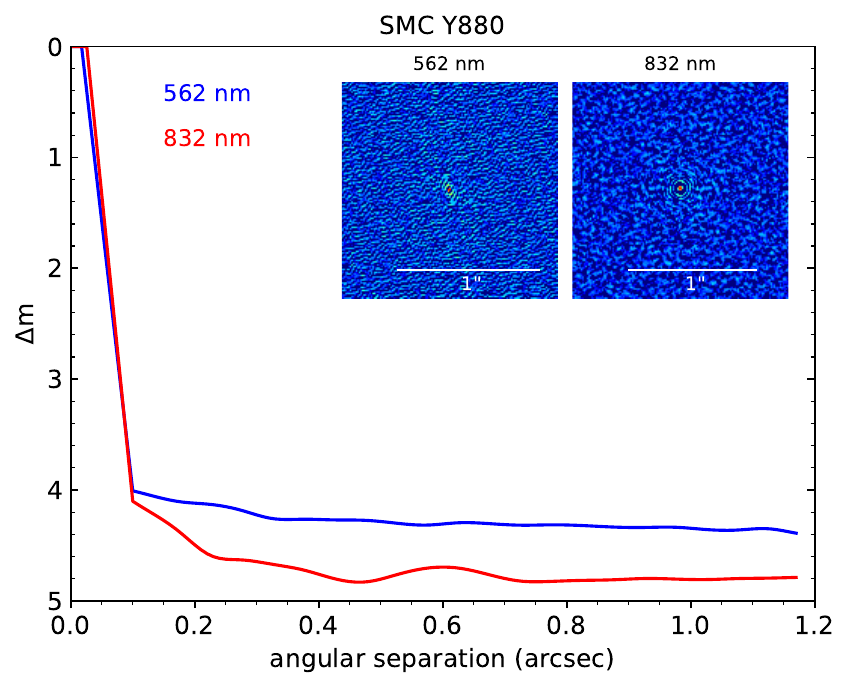}
\includegraphics[scale=0.4]{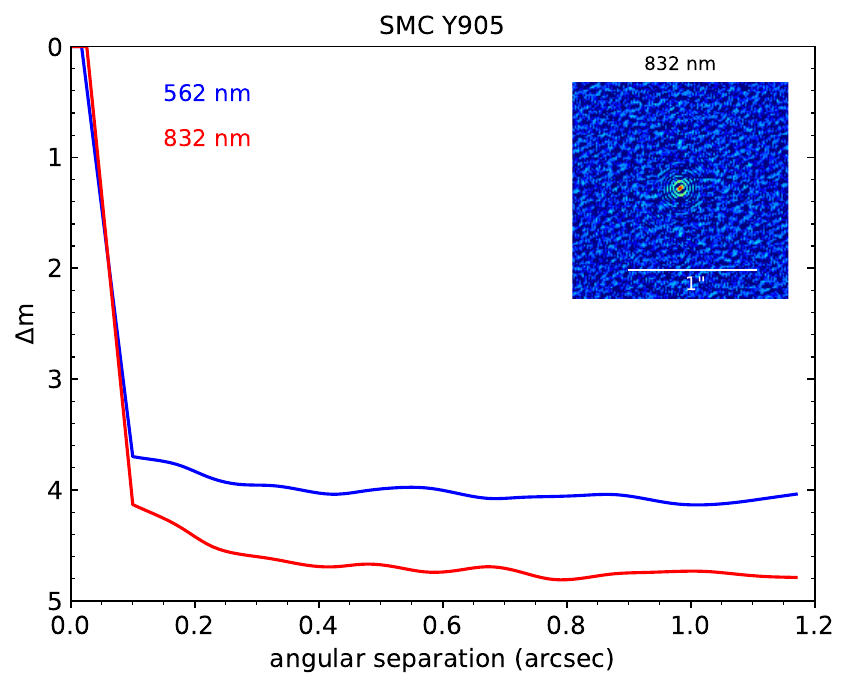}
\includegraphics[scale=0.4]{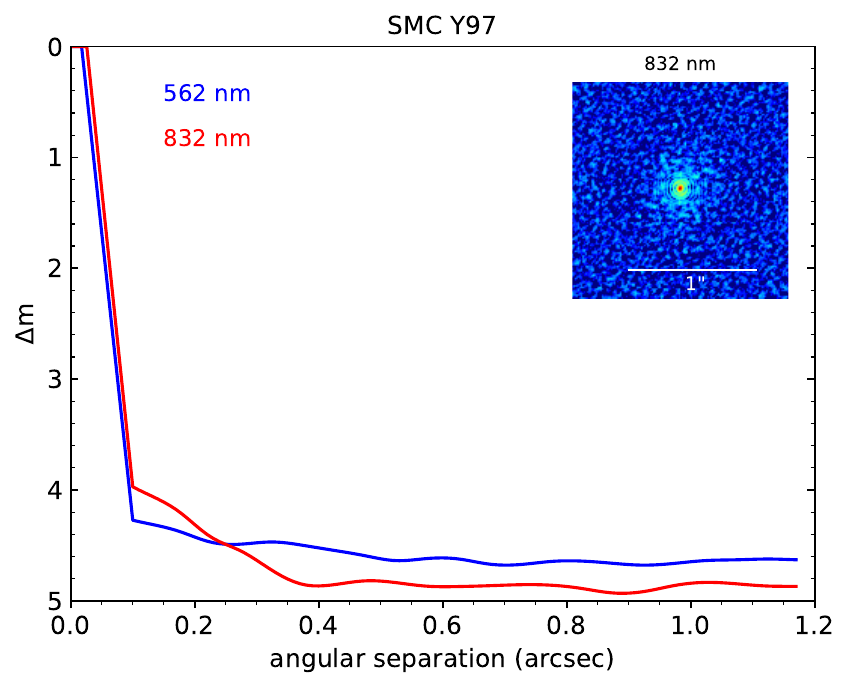}
\caption{Contrast curves, and reconstructed images of all the observed targets. Red (blue) denotes contrast curves in the 832nm (562nm) filters, while the reconstructed image in one or both of the filters are shown in the top right hand corner. The ordinate refers to the $\Delta$m in the respective filters. Object names are given top of each panel. \label{resultsallfunal}}

\end{figure}

\newpage

\bibliography{sample631}{}
\bibliographystyle{aasjournal}



\end{document}